\documentclass[twocolumn,preprintnumbers,amsmath,amssymb]{revtex4}

\usepackage{amsmath,amssymb,color,epsfig,latexsym,natbib,graphicx,dcolumn}
\usepackage{bm}% bold math
\bibpunct{[}{]}{,}{n}{,}{,} % see natbib.sty
\def\AD#1{{\textcolor{magenta}{#1}}} % plug a value, a ref, ...
\def\DEL#1{{\textcolor{green}{}}} % suggested deletion in text
\def\etal{et al.}

\def\ni{\noindent}

\def\rms{r.m.s.}

\newcommand{\bB}{\textbf{B}}

\newcommand{\bk}{\textbf{k}}
\newcommand{\bp}{\textbf {p}}
\newcommand{\bq}{\textbf {q}}

\newcommand{\bv}{\textbf{v}}

\newcommand{\bx}{\textbf{x}}

\newcommand{\rem}[1]{}

\newcommand\vecp[1]{\vec{#1}} % physical vector

\newcommand{\be}{\begin{equation}}
\newcommand{\ee}{\end{equation}}

\def\bB0{\vecp{B}_0}   

\begin{document}
\title{\bf Spectral Modeling of Rotating Turbulent Flows} %\\

\author{J. Baerenzung$^1$, P.D. Mininni$^{1,2}$, A. Pouquet$^1$,
    H. Politano$^{3,4}$, and Y. Ponty$^{3,4}$}
\affiliation{$1$ TNT/NCAR, P.O. Box 3000, Boulder, Colorado 80307-3000,
    U.S.A. \\
$2$ Departamento de F\'\i sica, Facultad de Ciencias Exactas y
    Naturales, Universidad de Buenos Aires, Ciudad Universitaria, 1428
    Buenos Aires, Argentina. \\
$3$ Laboratoire Cassiop\'ee, UMR 6202, Observatoire de la C\^ote d'Azur,
    B.P. 4229, 06304 Nice Cedex 4, France. \\
$4$ Universit\'e de Nice-Sophia Antipolis, CNRS UMR 6202, Observatoire
    de la C\^ote d'Azur, B.P. 4229, 06304 Nice Cedex 4, France.}

\begin{abstract}
We test a subgrid-scale spectral model of rotating turbulent flows against direct numerical simulations.
The particular case of Taylor-Green forcing at large scale is considered, a configuration that mimics the flow
between two counter rotating disks as often used in the laboratory. We perform computations in the presence of
moderate rotation down to Rossby numbers of 0.03, as can be encountered in the Earth atmosphere and oceans. We
provide several classical measures of the degree of anisotropy of the small scales of the flows under study
and conclude that an isotropic model may suffice at moderate Rossby numbers. The model, developed previously
(Baerenzung et al., Phys. Rev. E {\bf 77}, 046303 (2008)), incorporates eddy viscosity that depends
dynamically on the inertial index of the energy spectrum, as well as eddy noise. We show that the model
reproduces satisfactorily all large-scale properties of the direct numerical simulations up to
Reynolds numbers of $\sim 10^4$ and for long times after the onset of the inverse cascade of energy at low Rossby number.
\end{abstract}
\pacs{}
\maketitle

%%%%%%%%%%%%%%%%%%%%%%%%%%%%%%%%%%%%%%%%%%%%%%%%
\section{Introduction}
Rotating flows are commonplace in nature, the influence of rotation being measured by the Rossby number $Ro=U_0/2L_0\Omega$,
with $U_0$ the r.m.s. velocity, $L_0$ a characteristic lengthscale of the flow and $\Omega$ the rotation rate.
The Rossby number of the atmosphere is $\sim 0.1$ and in the ocean it can be as small as $10^{-3}$.
Assuming a constant rotation rate, the Coriolis force that appears in the equations leads to the emergence
of wave motions which, at small enough Rossby number, can be thought as dominating the dynamics.
However, at high Reynolds number, $Re=U_0L_0/\nu$, with $\nu$ the viscosity, turbulent eddies interact
with waves, and inertial waves interact nonlinearly (in particular through resonances), so that the dynamics become complex.
The Rossby number can be viewed in this way as the ratio of the characteristic time of an inertial wave,
$\tau_{W} \sim 1/\Omega$ to the characteristic time of an eddy, or eddy turn-over time, $\tau_{NL}\sim L_0/U_0$;
when small, the waves are rapid and may dominate the dynamics.
%, at least for short times (of order $1/\Omega$).

Many studies have been devoted to the exploration of rotating turbulence, experimental as well as numerical and theoretical 
(see e.g.\cite{cambon_book}).
One expects the flow to become quasi bi-dimensional under the influence of strong rotation but recent studies show that
the dynamics is more subtle, with three-dimensional eddies possibly prevailing at small scales.
The case of small Rossby number can be studied using anisotropic extensions of closure models, such as
the Eddy Damped Quasi Normal Markovian approximation (EDQNM hereafter).
In such approaches, the closure is obtained by modeling the damping of fourth-order cumulants
(non-zero for a non Gaussian field) by a term linear in third-order moments; dimensionally, the constant
of proportionality is the inverse of a time, or a rate $\mu$, taken in EDQNM to be the rates known to be significant
in the physics of the problem. In the simplest case of non-rotating isotropic and homogeneous turbulence,
these rates are proportional to the inverse of the eddy turn over time $\tau_{\ell}=\ell/U_{\ell}$ and of the viscous time $\tau_D\sim \ell^2/\nu$,
expressed in terms of the scale $\ell$ and the velocity at that scale $U_{\ell}$. In the rotating case, the wave frequency
becomes relevant as well \cite{cambon_e1} (see \cite{holloway} for an early realization of this concept), and because of
the anisotropic dispersion relation of inertial waves, the model becomes anisotropic itself in terms of a spectral energy
distribution that is a function of the wavenumbers $k_{\perp}$ and $k_{\parallel}$ \AD{}, where $\perp$ and $\parallel$ refer
to directions relative to the rotation axis.

Other approaches include weak turbulence theory \cite{galtier03}, following the original methodology of Benney and Newell \cite{benney},
and other resonant wave theories \cite{Greenspan,Waleffe93} recently shown to correspond to an
asymptotic limit for flows with wave dynamics \cite{Embid96}.

The link between resonant theories and the EDQNM closure for rotating flows has been analyzed in detail recently
(\cite{cambonNJP06, belletJFM08} and references therein); it can be simply said here that when the global damping rate
$\mu$ (omitting triadic scale dependence) is dominated by the inertial frequency, in the limit of
$\tau_{W} \rightarrow 0$,
the fourth-order cumulant becomes negligible, the flow becomes quasi-gaussian and the closure
occurs naturally. Of course, this limit needs to be taken carefully, in particular when approaching the slow manifold
($k_{\parallel}=0$) \cite{smith99,cambonNJP06}, possibly because of what could be called interferences between
resonant modes and modes in the slow manifold.

The Coriolis force does not affect the kinetic energy balance nor does it modify the nonlinear part of the exact
law stemming from energy conservation \cite{PP98,Lindborg95} when stated in its anisotropic version. Similarly, helicity
conservation is not altered by the Coriolis force, nor is (in the structure function formulation) the nonlinear part of the exact law
stemming from that conservation \cite{PPG00}: indeed, helicity is conserved in the presence of rigid body rotation and, when using
structure functions as in \cite{PPG00}, the constant rotation vector drops from the dynamical equation for the second order moment. 
The fact that uniform rotation affects odd-order
moments of the velocity but not even ones (at least in the linear limit of negligible nonlinear terms) shows that its effects
are subtle. However, it has been documented in the literature (see e.g. \cite{bardina}) that a specific model is needed in
Large Eddy Simulations (LES) of turbulent rotating flows in order to take into account the slowing down of energy decay because of
waves \cite{davidson}, and the anisotropy of integral length scales and of dissipation.

Furthermore, wave resonant theories are non uniform in scale and it is not clear whether their predictions are verified in high Reynolds number flows in the laboratory, in the environment or in direct numerical simulations (DNS).
In the decaying case \cite{jacquin90}, there may be different temporal regimes in which different mechanisms prevail \cite{belletJFM08}. In a forced flow, high resolutions and long-time integrations are needed in order to resolve the different spatial regimes that may develop (e.g. inverse and direct cascades of energy \cite{Bardina85,Bartello94,Smith96,Cambon97,us_alex} and direct cascades of helicity \cite{us_helical}), as well as the short-time wave regime versus the long-time turbulent regime.

Modeling becomes a necessity in order to reach an understanding of these flows at high Reynolds numbers as occurs in astrophysics and geophysics. Using a closure set of integro-differential equations for the energy spectra (see \cite{belletJFM08} and references therein) or the weak turbulence framework developed in \cite{galtier03} is a powerful tool
 for sufficiently small Rossby number, but two difficulties have to be overcome. On the one hand, such techniques are valid in the limit $Ro\rightarrow 0$ and yet the flows one tries to model may present inhomogeneities (in space, in time, in scale) that are affected differently by the rotation. On the other hand, the complexity of the so-called weak turbulence kinetic equations and in particular their dependence on the angle between the wavevector (in a Fourier decomposition) and the rotation axis (to be taken as the z axis in what follows) necessitates a regular discretization in angle as opposed to an exponential discretization in wavenumber, the latter working because of the self-similarity of the known (power law) spectral solutions to the equations. This angular dependency makes the closure or weak turbulence equations difficult and costly to use (see however \cite{belletJFM08}).

The anisotropy of a rotating flow comes from the nonlinear terms (and resonant triadic interactions) \cite{davidson}, at least for one-time second order statistics \cite{gence}, because of a loss of phase information. An explicit example of an initial condition that gives rise to elongated vortices is given in \cite{davidson}; these authors predict linear growth of the integral scale, as opposed to the classical Kolmogorov $\sim t^{2/7}$ law, and as observed in laboratory experiments \cite{jacquin90}. This anisotropy is also linked to an asymmetry (predominance of cyclonic event over anti-cyclonic) for times of order $1/\Omega$, as measured for example by the skewness of the vertical component of the vorticity \cite{morize,moisy08}.
However, it has been shown by several authors that the expected bi-dimensionalization of the flow is only realized partially, and small scale eddies may not follow such a dynamics; in which case, one expects the small scale eddies (i.e., those that are to be modeled in an LES approach) to be somewhat isotropic. It may thus be envisageable
to use, as a model of small scales, a methodology developed for isotropic flows.

It is in this context that we extend the spectral model derived in \cite{LESPH} to the case of forced rotating flows, comparing the results of the model to high resolution DNS \cite{us_alex} %us_helical}
for forced rotating turbulence down to $Ro\sim 0.03$. The model is based on the EDQNM closure to compute eddy viscosity and eddy noise. It adapts dynamically to the inertial index of the energy spectrum, and as a result it is well suited to study rotating turbulence for which the scaling laws are not well known, and may change with the Rossby number, and also (at fixed Rossby number) as the system evolves and an inverse cascade develops. The next section poses the problem in terms of equations and models and gives the numerical set-up. We then describe the results for the isotropic LES model, examining energetic balance, structures, spectra and higher-order statistics. Finally, the last section presents our conclusions.

\section{Equations and spectral modeling}\label{s:modeling}
\subsection{Primitive equations}\label{ss:modeling}

 The dynamical equations can be written in terms of the Fourier coefficients of the
velocity field defined as usual as:
\begin{equation}
\textbf{v}(\textbf{k},t) = \int\!\!\! \int\!\!\! \int_{-\infty}^\infty \textbf{v}(\textbf{x},t)
e^{-i\textbf{k}.\textbf{x}} \textbf{dx} \ .
\end{equation}
In the rotating frame, and including the centrifugal force in the pressure term, the equations are:
\begin{equation}
\left(\frac{\partial}{\partial t}+\nu k^2\right)v_\alpha(\bk,t) +2\Omega P_{\alpha\beta}\varepsilon_{\beta z \gamma}u_\gamma(\bk,t) =
{t_\alpha}(\bk,t) +{F_\alpha}({\bf k})
\label{mhd_eq1} \end{equation}
\ni together with the incompressibility condition ${\bk \cdot \bv} = {\bf 0}$;
 $\nu$ is the kinematic viscosity, ${\bf F}({\bf k})$ is the Fourier transform of the forcing function,
$P_{\alpha\beta}=\delta_{\alpha\beta}-k_\alpha k_\beta/k^2$ is the projection operator, $\Omega$ is the rotation rate and
$\textbf{t}(\bk,t)$ is a bilinear operator for the kinetic energy transfer written as:
\begin{equation}
t_\alpha (\bk,t) =-iP_{\alpha\beta}(\bk)k_\gamma \sum_{\textbf{p}+ \textbf{q} = \textbf{k}} v_\beta(\bp,t) v_\gamma(\bq,t) \ .
\label{mhd_eq2}
\end{equation}
%\NOTE{Shouldn't the projector operator also be applied to the Corilis term? It is not solenoidal unless the projection is done, because $\nabla \cdot (\Omega \times \bv) = - \Omega \cdot (\nabla \times \bv)$.} where $P_{\alpha\beta}(\bk)=\delta_{\alpha\beta} - {k_\alpha k_\beta}/{k^2}$
Note that $P_{\alpha\beta}$ is a projector that allows us to take the pressure term of the velocity equation
into account via a Poisson formulation
and ensures that the velocity remains divergence-free including in the presence of rotation.
Finally note that the total energy
$E_T=\left<{\bf v}^2/2\right>$
and the helicity $\left<{\bf v} \cdot {\bf \omega}\right>$ (with ${\bf \omega}=\nabla \times {\bf v}$) are invariants of
the three-dimensional equations in the ideal case, i.e. in the absence of viscous dissipation ($\nu=0$).
Besides the Reynolds number and the Rossby number defined previously, one can also introduce dimensionless numbers based
on small-scales as produced by the turbulent flow; the simplest way to do that, traditionally, is to base such parameters
on the vorticity through the Taylor scale $\lambda$ defined as:
\begin{equation}
\lambda = 2\pi \left(\frac{\int{E(k) dk}}{\int{E(k) k^2 dk}}\right)^{1/2};
\label{eq:taylor}
\end{equation}
the Taylor Reynolds number is then:
$$R_{\lambda}= \frac{U_0 \lambda}{\nu} \ .$$
One can also define a quantity called the micro-Rossby number \cite{Cambon97} which is useful to
determine the regime of the small scale turbulence and the slope of the energy spectrum \cite{morize}. It reads:
\begin{equation}
Ro_\omega = \frac{\omega_{rms}}{2\Omega},
\label{mRo}
\end{equation}
where $\omega_{rms}$ stands for the $\rms$ vorticity; note that it is proportional to the Rossby number evaluated at the Taylor scale.
Finally, we also define the Ekman number:
\begin{equation}
Ek = \frac{Ro}{Re} = \frac{\nu}{2 \Omega L_0^2} ,
\end{equation}
where $L_0=2\pi/k_0$ is the scale associated with the forcing at $k_0=2$.
The direct numerical simulations used in this paper (runs Id, IId and IIId respectively, see Table \ref{table1})
in order to assess the validity of the LES are those labeled A3, A4 and A6 respectively in \cite{us_alex} (hereafter, paper I).
For all these runs, the forcing function is a Taylor-Green (TG) vortex with amplitude $F_0$:

\begin{eqnarray}
{F_x}&=&F_0 \sin (k_0 x)\cos(k_0 y) \cos(k_0 z)\nonumber \\
{F_y}&=&-F_0 \cos (k_0 x)\sin(k_0 y) \cos(k_0 z)\nonumber \\
{F_z}&=& 0 \ ;
\label{TG2}
\end{eqnarray}
the third component of the forcing is equal to zero but the velocity in the z-direction grows through nonlinear interactions.
Moreover, the forcing injects no energy in modes with $k_z=0$, and as a result any amplification observed in strongly rotating cases
must be only due to a cascade process. Finally, the resulting flow has a small spectral anisotropy with slightly more energy in
the $z$ direction \cite{us_alex}, an effect which is the opposite of the tendency towards two-dimensionalization that develops in rotating turbulence.

The numerical computations using the above forcing are thus either Direct Numerical Simulations of the Navier-Stokes
equations with $256^3$ grid points, or Large Eddy Simulations on grids of $64^3$ points; the axis of rotation is the $z$-axis,
and the flow is initially at rest. Note that the TG flow is widely used in experimental devices to study turbulence and its effect
on the generation of magnetic fields \cite{pinton} even though the TG vortex has no net helicity due to its symmetries; because of
this latter property, the LES model used here will not include the helicity eddy viscosity derived in \cite{LESPH} (Paper II hereafter). The turnover time
at the forcing scale is then defined as $\tau_{NL} = L_0/U_0$ where $U_0=\sqrt{\left< {\bf v}^2 \right>}$ is the r.m.s. velocity
measured in the turbulent steady state as stated previously, at the onset of the inverse cascade at low Rossby number.
Note that the amplitude of the forcing $F_0$ in each simulation is increased with $\Omega$ to have $U_0 \approx 1$ in all the runs
before the inverse cascade sets in (see \cite{us_alex} for more details on the DNS runs).

Finally, as the issue of the direction of the energy cascade (direct and/or inverse) is an important issue in rotating turbulence,
a useful diagnostic in this context is to examine the behavior of the skewness (normalized third-order moment corresponding to energy transfer)
based on the velocity derivative, %evaluated on a distance $r$, % $\partial v(r)$, or on the vertical vorticity $\omega_z$ (see e.g. \cite{Bartello94}), namely:
\be
S_k= {  \frac{\left<(\partial_x v_x)^3\right>}{\left<(\partial_x v_x)^2\right>^{3/2}} } \ .
\label{SKEW} \end{equation}
%\ \ \ S_{\omega}= \frac{ <\omega_z^3>}{<\omega_z^2>^{3/2}}\ . $$

\subsection{The isotropic EDQNM closure}
The Large Eddy Simulation model (LES) derived in \cite{LESPH} for non-rotating Navier-Stokes flows is now extended
to the rotating case in its non-helical version (LES-P of Paper II). In other words, intrinsic variations of the
helicity spectra are not taken into account in the present work in the evaluation of the transport coefficients used in our LES model.
The first step of the model is to employ a spectral filtering of the
equations; this operation consists in truncating
all velocity components at wave vectors $\bk$ such
that $|\bk | = k > k_c$, where $k_c$ is a so-called cutoff wave number.
Since the scales associated with $k_c$ are presumably much larger than the
%the cutoff are larger than the
actual dissipative small scales in a high Reynolds number flow, one needs to model the transfer between the
large (resolved) scales and the small (subgrid unresolved) scales of the flow.
In order to approximate these transfer terms,
the behavior of the energy spectrum after the cutoff wave number has
to be estimated. We therefore define an intermediate range, lying between $k_c$ and $3k_c$,
where the energy spectrum is assumed to present a power-law behavior possibly
followed by an exponential decrease:
\begin{equation}
E^V(k,t) = E_0^V k^{-\alpha_E^V} e^{-\delta_E^V k}, \quad k_c\le k<3k_c \ . \label{fit_Ev}
\end{equation}
The coefficients $\alpha_E^{V}$, $\delta_E^{V}$ and $E_0^{V}$ %, and $\alpha_E^M$, $\delta_E^M$, $E_0^M$,
are computed at each time step,
through a mean square fit of the resolved energy spectrum.
In a second step, one can write the following model equations (omitting forcing):
\begin{eqnarray}
\left[\partial_t + \left(\nu\left(k|k_c,t\right) +\nu \right)k^2 \right]
v_\alpha(\textbf{k},t) & & \nonumber \\
=t_\alpha^{<}(\bk,t) - 2\Omega P_{\alpha\beta}\epsilon_{\beta z \gamma}u_\gamma(\bk,t) &&\ ,
\label{mhdmodel}
\end{eqnarray}
where the $<$ symbol indicates that the nonlinear transfer terms are
integrated over a truncated domain defined such that $\bp +\bq = \bk$ with $|\bp|=p, |\bq|=q < k_c$. The eddy viscosity
$\nu\left(k|k_c,t\right)$ is expressed as
\begin{equation}
\nu(k|k_c,t)\!\!\! \ = \!\!\! \ - \iint_{\Delta^>}\!\!\!\theta_{_{kpq}}
\frac{S_{E_2}(k,p,q,t)}{2 k^2E^V(k,t)} \ dpdq \ . \nonumber
\end{equation}

The function $S_{E_2}(k,p,q,t)$ corresponds to the so-called
absorption term (linear in the energy spectrum $E^{V}(k,t)$) in the EDQNM nonlinear transfer, lending itself
in particular to an expression for the turbulent eddy viscosity, as is well known;
$\Delta^>$ is the integration domain over (${\bf k}$, ${\bf p}$, ${\bf q}$) triangles,
such that $p$ and/or $q$ are larger than $k_c$, and both $p$ and $q$ are
smaller than $3k_c$.

Finally, to take into account the effect of the emission (eddy-noise) term involved in the EDQNM nonlinear
transfer ($S_{E_1}(k,p,q,t)$), we use a reconstruction field procedure which enables us to partly rebuild the phase
relationships between the three spectral components of the velocity field, as explained in detail in Paper II \cite{LESPH}.
The functions $S_{E_1}(k,p,q,t)$ and $S_{E_2}(k,p,q,t)$ appearing in the transport coefficients used in the LES are written for completion in the Appendix.
Note that, although isotropic, the subgrid model we use in this paper has an important feature: it adjusts dynamically to
the energy spectrum instead of assuming a given spectral law, usually the classical Kolmogorov law, $E(k)\sim k^{-5/3}$.
This allows for exploration of flows for which a theory to predict spectral indices is not available. Also note that the
reconstruction procedure differs as well from traditional implementations insofar as it tries to keep the phase information of the small-scales.

%\NOTE{I think more should be said here on how timescales are adjusted to ``adapt dynamically'' to the spectral index of the flow. For someone that hasn't read Paper I (II?), this looks like an isotropic version of what Cambon did.}

%%%%%%%%%%%%%%%%%%%%%%%%%%%%%%%%%%%%%%%%%%%%%%%%%

\section{Rotation and isotropy}
\label{rot_an}

One of the effects of rotation on a flow is to induce anisotropy, as in the formation of large-scale columnar vortices.
In that light, we explore in this section the anisotropic properties of a DNS at low Rossby number to see whether or not it is
 relevant to use a model based on isotropic assumptions
to simulate a flow subjected to rotation.
The LES model we propose to use approximates, as is customary, the transfer from the large
to the small scales, but most of the modeled interactions are
between small scales because of the value of $k_c$ (chosen to be in all cases larger than the energy injection wavenumber),
and because most of the modes in a turbulent flow are in the small scales
(recall that the number of modes in a given isotropic shell $k_i$ varies as $k_i^2$).

We therefore investigate now the
properties of the small scales of flows forced with the Taylor-Green vortex (see Eq. \ref{TG2}) and subjected to rotation,
with $k_0=2$ and at a Rossby number $Ro=0.03$; we perform a DNS on a grid of
$256^3$ points and with the flow being initially at rest.
To measure anisotropy, we introduce two different quantities, a spatial one
and a spectral one, denoted respectively $I^D$ (for dimensional) and $I^C$ (for Craya \cite{craya}; see also \cite{Herring}).
Another measure of anisotropy linked to the so-called polarization anisotropy, following \cite{cambon_jacquin}
(see also \cite{morinishi}), is discussed later in Section \ref{aniso}.
% (see Fig. \ref{b33}).
The spatial coefficient $I^D$ evaluates
the averaged ratio between the intensity of the velocity in the
perpendicular direction $V_\perp(\bx,t)$ and in
the parallel direction $V_\parallel(\bx,t)$, with $\perp,\ \parallel$ referring to the $z$-axis of rotation. The velocity field can
be expressed as a function of these two components as
${\textbf v} (\bx,t) = V_\parallel(\bx,t)
{\textbf e_\parallel} + {\bf V_\perp}(\bx,t)$, where
${\textbf e_\parallel}$ is the unit vector associated to the axis of rotation
and ${\textbf V}_\perp(\bx,t)$ is the velocity field projected on the plane perpendicular to ${\textbf e_\parallel}$.

%${\textbf e_\perp} = \frac{1}{x}\bx \cdot {\textbf e_x} + \frac{1}{x}\bx \cdot {\textbf e_y}$. \NOTE{I don't understand this notation. The way ${\textbf e_\perp}$ is written here, it is a scalar. Perhaps you want to say ``...${\textbf v} (\bx,t) = V_\parallel(\bx,t) {\textbf e_\parallel} + {\textbf V}_\perp(\bx,t)$, where ${\textbf e_\parallel}$ is the unit vector associated to the axis of rotation and ${\textbf V}_\perp(\bx,t)$ is a vector field perpendicular to ${\textbf e_\parallel}$.
The spatial anisotropy coefficient therefore reads:
\begin{equation}
I^D = \left< \frac{V_\perp(\bx,t)}{V_\parallel(\bx,t)}\right> \ .
\label{icr} \end{equation}
The spectral coefficient $I^C$ is computed as in \cite{curry}:
for each wavevector ${\bf k}$, an orthonormal reference frame is defined as
(${\bf k}/|{\bf k}|$, ${\bf e}_1({\bf k})/|{\bf e}_1({\bf k})|$,
${\bf e}_2({\bf k})/|{\bf e}_2({\bf k})|$), with ${\bf e}_1({\bf k})=
{\bf k} \times {\bf z}$ and ${\bf e}_2({\bf k})={\bf k} \times
{\bf e}_1({\bf k})$, where ${\bf z}$ is the vertical unit wavevector.
In that frame, since the incompressibility condition yields
${\bf k} \cdot {\bf v}({\bf k}) = 0$, ${\bf v}({\bf k})$ is only determined
by its two components ${\bf v}_1({\bf k})$ and ${\bf v}_2({\bf k})$.
This second anisotropy coefficient is then defined as
\begin{equation}
I^C=\sqrt{\left<|{\bf v}_1({\bf k})|^2\right>/\left<|{\bf v}_2({\bf k})|^2\right>} \ .
\label{ici} \end{equation}
Both $I^D$ and $I^C$ are such that they have unit values for fully isotropic flows.

\begin{figure}
\includegraphics[width=\linewidth,height=4cm]{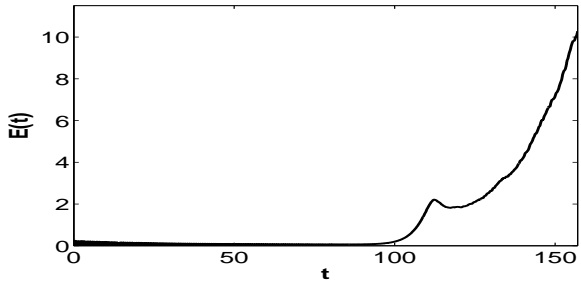}
\caption{Temporal evolution of the energy for the flow in a DNS with $Ro=0.03$; it displays two temporal phases, one dominated by wave interactions and one, for t larger
than $100$, corresponding to an inverse cascade of energy.}
\label{evo_energy_dns} \end{figure}

In Fig. \ref{evo_energy_dns} we plot the temporal evolution of the total energy, the time being expressed in units of
the eddy turn-over time. Note the long interval before turbulence fully develops, as rotation is strong and the run was
started from a fluid at rest.
%(note that this evolution is slightly different from that in Paper I, due to slightly different initial conditions.
Indeed, before the energy starts to grow at $t\approx 90$, one can observe a long transient during which the energy displays
damped oscillations in time (see Paper I). This transient is linked to the effect of rotation and its duration increases
linearly with $\Omega$, i.e. as the inverse of the Rossby number. During this first stage, the energy dissipation rate
is small and the energy spectrum is very steep. Later, at $t\approx 90$, the enstrophy starts to grow and the energy dissipation
rate increases. The energy also grows and an inverse cascade of energy develops.
Turbulence sets in and the small-scale energy spectrum develops an inertial range with scaling close to $\sim k_\perp^{-2}$ (see Paper I for more details).

In order to quantify the importance of anisotropy at what would be the sub-grid scales in a LES of rotating flows,
we start by noting that the velocity (in particular when an inverse cascade of energy develops at small enough Rossby number) is
dominated by the large scales whereas the modeling will occur in the small scales of the flow. In this context, we introduce
a band-pass filter of the DNS data in order to concentrate the analysis on small-scale properties of the flow. The filtered field
%, which actually corresponds
%in real space to the small scales field of the flow,
is given in Fourier space by all the velocity components at a wavector $\bk$ such that $32 \leq |\bk | \leq 64$; note that, for
this DNS using a classical 2/3 dealiasing rule, the maximum wavenumber is $k_{max}=85$. As a result, the band-pass filter
can be interpreted as preserving the small scales of the direct cascade inertial range.

\begin{figure}
\includegraphics[width=\linewidth,height=4cm]{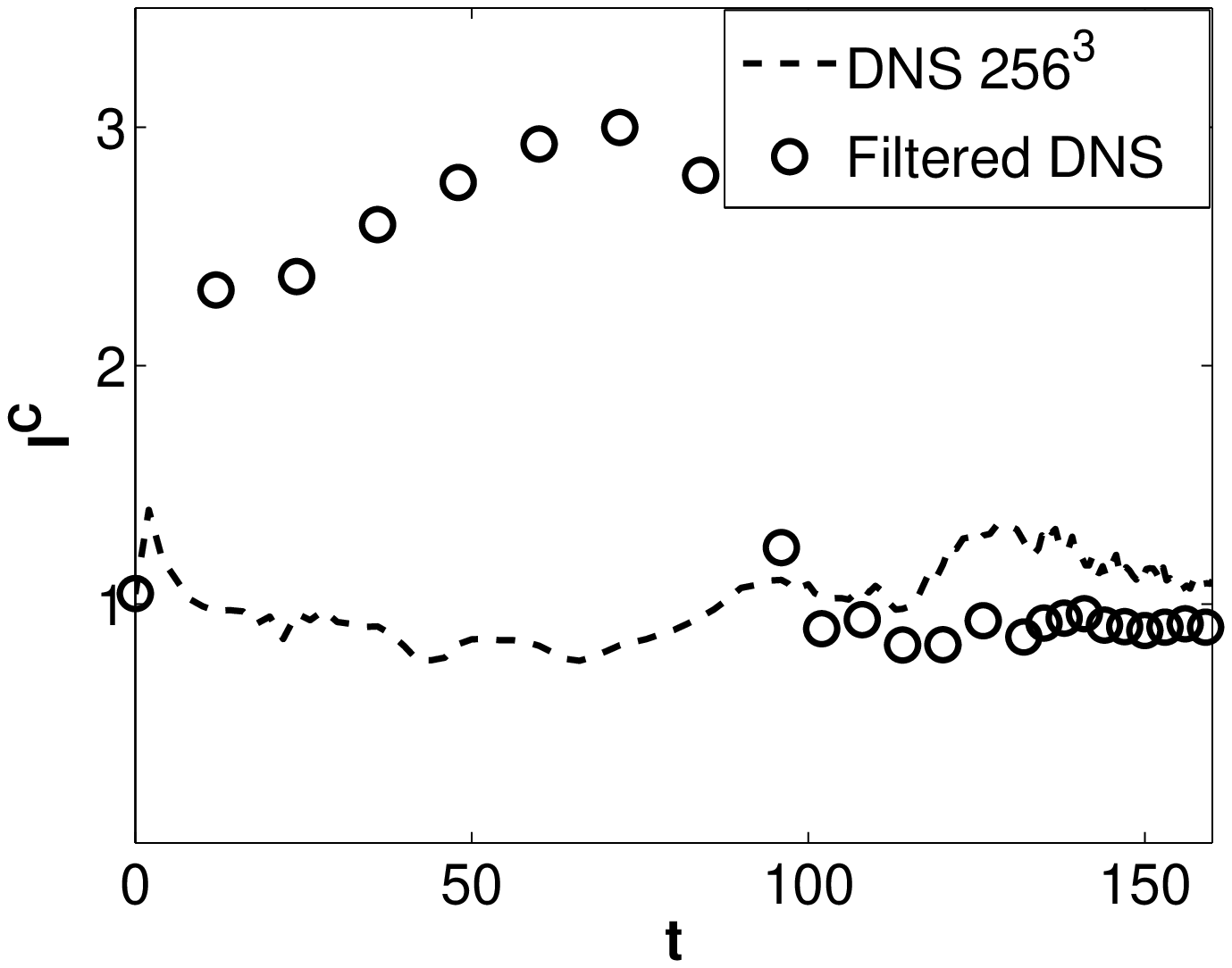}
\includegraphics[width=\linewidth,height=4cm]{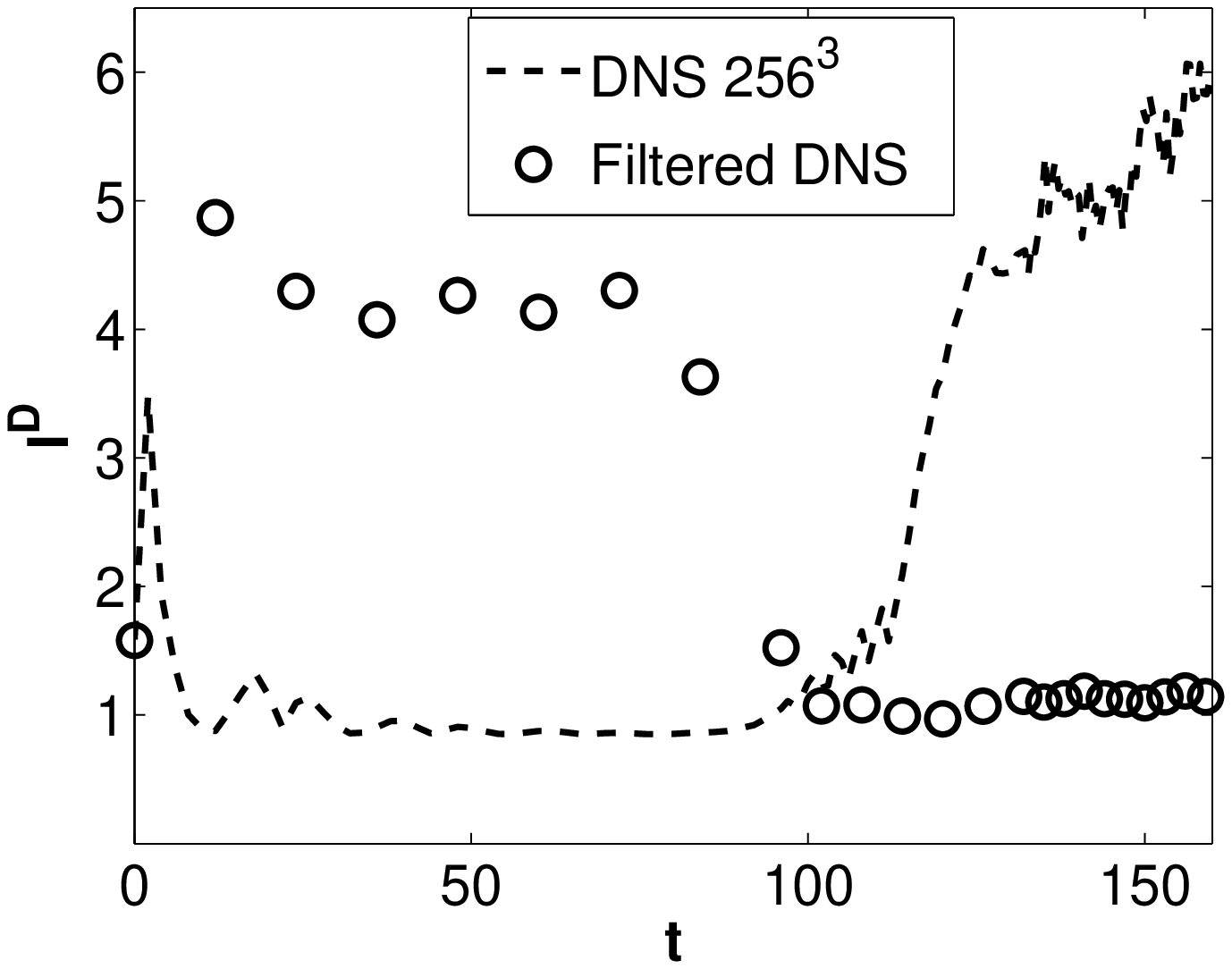}
\caption{Temporal evolution for a flow with $Ro=0.03$ of the Craya anisotropy coefficient $I^C$
(top) and the directional anisotropy coefficient $I^D$ (bottom) (see eqs. \ref{icr},\ \ref{ici}),
for the full DNS velocity field (dash line) and the filtered DNS field (ovals) defined as a band-pass filter for wavevectors in $|K|\in [32,64]$.
Note the sharp transition towards isotropy of the small scales for $t\simeq 100$, as both the direct and inverse turbulence cascades finally develop.}
\label{icr_ici}\end{figure}

Figure \ref{icr_ici} represents the time history of the $I^C$ and $I^D$
anisotropy coefficients for the complete DNS (dash line) and for
the band-pass filtered velocity fields of the flow at $Ro=0.03$ (ovals).
The Craya spectral coefficient $I^C$ of the complete DNS field remains close
to unity during the whole simulation, indicating that globally
the flow is close to an isotropic state. However, the directional coefficient
$I^D$ exhibits three different regimes in the full DNS: a first phase
between $t=0$ and $t\simeq 40$ during which it oscillates, with an
amplitude that decreases with time, and a second phase, between $t\simeq 40$
and $t\simeq 90$, with this coefficient remaining constant at a value close to unity,
meaning that no direction is privileged in the flow. Finally, in a third and last phase, which begins when the turbulence starts to develop, $I^D$
strongly increases with time. This behavior is the signature of the
generation of intense columnar structures within the flow,
within which the perpendicular component of the velocity field dominates
the parallel one.

The behavior of these coefficients is completely different for the filtered,
small-scale, field. Indeed, the small scales
are strongly anisotropic during the transient period before the turbulence
develops, with a maximum value for $I^C$ of the order of $3$ (and $5$ for $I^D$). In this phase,
the directional anisotropy coefficient clearly shows that the perpendicular
component of the velocity dominates the parallel one, and therefore
that the small scales are mostly bi-dimensional. At $t \simeq 80$, both
coefficients drop rather abruptly to a value of order unity, indicating that
when the turbulence develops the small scales become isotropic corresponding to a standard cascade of energy to small scales
(note that the scales for which the anisotropic and inverse accumulation of energy takes place are eliminated by the band-pass filter).

With this study of the small-scale behavior of a flow subjected to moderate rotation, we see that an isotropic LES model cannot
be used to treat every phase of the flow. Indeed, in the early transient phase, a model based on isotropic assumptions
will not be able to approximate properly the transfer between the subgrid scales and the resolved scales.
 We therefore decide to only use our model to study the turbulent regime of rotating flows, after $t\approx 100$ in the case 
of Figs. \ref {evo_energy_dns}, \ref{icr_ici}.
Moreover, this is consistent with the fact that a LES is designed to study turbulent flows, and cannot handle transitional
(laminar, wave-dominated) flows.
In the case of rotating flows starting from a fluid at rest, turbulence only develops after a transient time that depends linearly
with the magnitude of the rotation. Note that in many studies, simulations of rotating flows are started from a previous turbulent steady state,
and in that case our LES should have no problem to adapt as the spectral index changes with the evolution of the system.
%, as for example modeled in the phenomenology of \cite{Zhou95} describing a plausible spectral behavior for rotating flows (see \cite{galtier03} for the weak turbulence computation of the sepctrum).

Note that both coefficients $I^C$ and $I^D$ are relevant quantities in the context of this EDQNM-based LES: the
behavior of $I^C$ justifies the assumption of ``spectral isotropy'' (i.e., dealing with $k$ instead of $(k_{||},k_\perp)$ at small scales);
on the other hand, the behavior of $I^D$ justifies the isotropic reconstruction done with the eddy-noise,
because $I^D\approx 1$ is a measure of variance isotropy.

\section{Numerical tests of the LES} \label{first}

We now test our LES model against direct numerical simulations
with different Rossby numbers. As stated before, the forcing used is the Taylor-Green vortex (see Eq. \ref{TG2}) at $k_0=2$.
For each simulation, we follow the numerical procedure described in Paper I; namely, we vary the rotation
rate leading to three different Rossby numbers: $0.03$, $0.17$, and $0.35$.
The simulation parameters are summarized in Table \ref{table1}.
The flow evolves in a periodic box, with $256^3$ grid points for the DNS
and $64^3$ grid points for the LES. The ``reduced-DNS'' results, in the table and figures,
refer to the filtered DNS data on a grid of $64^3$ grid-points, corresponding to the limited information contained in the LES grid.
Since we are interested in studying only the modeling of the turbulent regime,
we start the LES simulations from the reduced-DNS data at a time after the end of the transient phase. However, if the LES is
started from a fluid at rest (i.e., started like the DNS at $t=0$), no significant differences are observed
with the procedure of starting the LES at the end of the transient phase,
except that the transient regime in the flow with $Ro=0.03$ is shorter. This accelerated evolution of the LES at low Rossby
number during the transient when compared to the DNS can be easily explained considering the inclusion of transport coefficients in the LES
which assumes that a turbulent flow is already present.

\begin{table*}
\caption{\label{table1}Parameters of the simulations:
Resolution $N^3$, Rossby number $Ro$ based on the forcing scale $L_0$,
Taylor microscale $\lambda$ and integral scale $L$,
\rms velocity $U_0=\left<{\bf v}^2\right>^{1/2}$,
integral Reynolds number $Re=$$\ U_0 L/\nu$ and eddy turnover time $\tau_{NL}=L_0/U_0$; $t_m$ is the final time of the computation.
Note that the {\bf r} label in the nomenclature of the runs stands for reduced data obtained by filtering in spectral space to $64^3$ points the original $256^3$ DNS data
$\lambda$, $L$, $Re$ and $\tau_{NL}$ are evaluated at the final time of the simulation for runs {\bf I}
which undergoes an inverse cascade, whereas they are averaged during the stationary phase of simulations {\bf II} and {\bf III} 
which are at higher Rossby numbers and do not undergo any significant inverse energy transfer.
}
\begin{ruledtabular}
\begin{tabular}{cccccccccc}
& & $N$ & $Ro$ & ${\lambda}$& $L$ &$U_0$& $Re$ & $\tau_{NL}$ & $t_m$\\
\hline
{\bf Id} & DNS & $256$ & $0.03$ & $2.06$ & $5.71$ & $4.53$ & $12920$ & $1.26$ & $157$ \\
{\bf Ir} & Reduced-DNS & $64$ & -- & $2.37$ & $5.71$ & $4.53$ & $12927$ & $1.26$ & -- \\
{\bf IL} & LES & $64$ & -- & $2.07$ & $5.59$ & $4.60$ & $12857$ & $1.22$ & -- \\
\hline
{\bf IId} & DNS & $256$ & $0.17$ & $0.65$ & $1.44$ & $1.01$ & $729$ & $1.41$ & $45$ \\
{\bf IIr}& Reduced-DNS & $64$ & -- & $0.73$ & $1.45$ & $1.01$ & $732$ & $1.44$ & --\\
{\bf IIL} & LES & $64$ & -- & $0.76$ & $1.49$ & $1.09$ & $813$ & $1.36$ &-- \\
\hline
{\bf IIId} & DNS & $256$ & $0.35$ & $0.77$ & $1.47$ & $1.07$ & $786$ & $1.36$ & $45$\\
{\bf IIIr}& Reduced-DNS & $64$ & -- & $0.72$ & $1.41$ & $0.96$ & $678$ & $1.46$ & -- \\
{\bf IIIL}& LES & $64$ & -- & $0.75$ & $1.42$ & $0.98$ & $695$ & $1.45$ & -- \\
\end{tabular} \end{ruledtabular} \end{table*}

\subsection{Global behavior of the flow} \label{global}
\begin{figure}
\includegraphics[width=\linewidth,height=4cm]{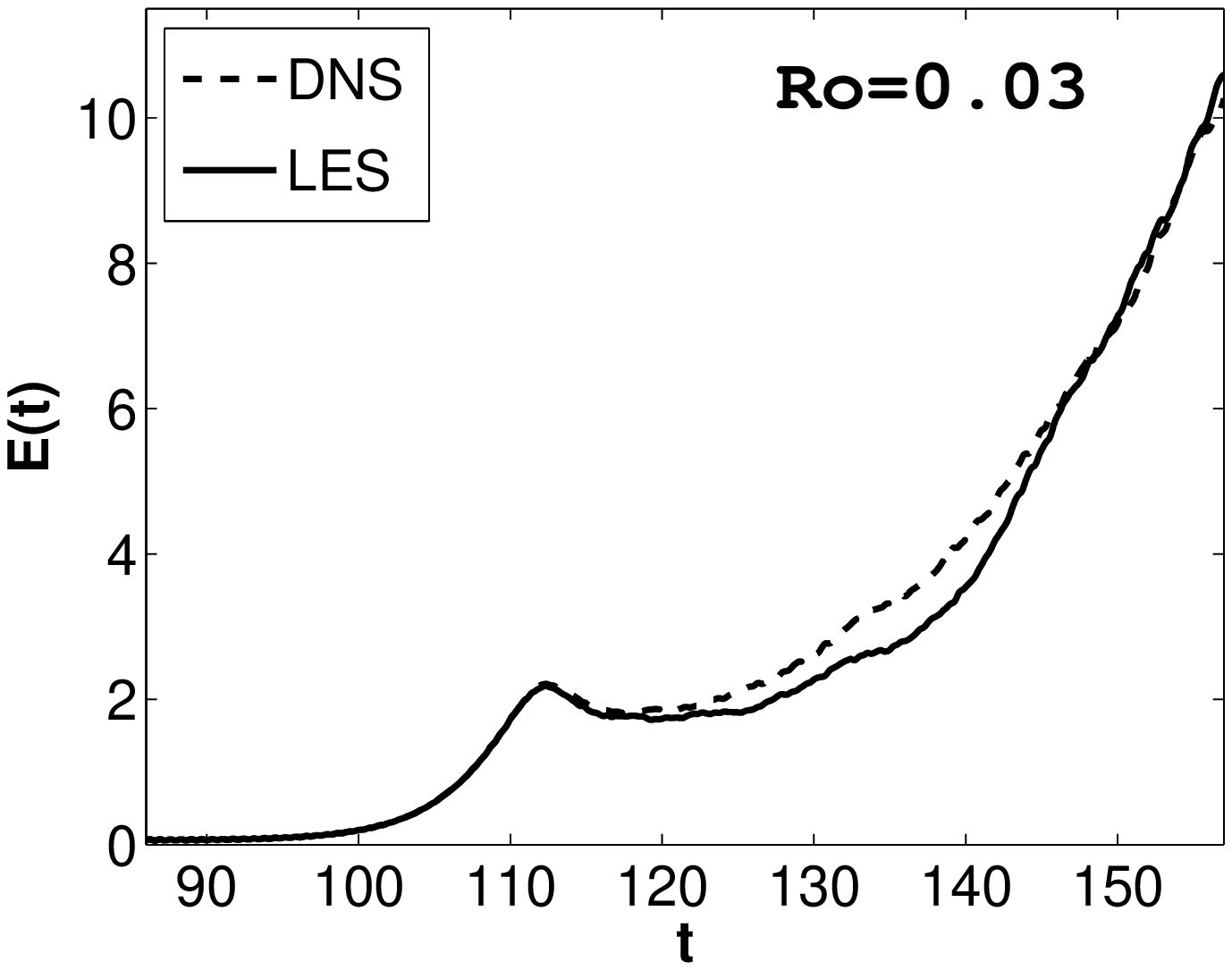}
\includegraphics[width=\linewidth,height=4cm]{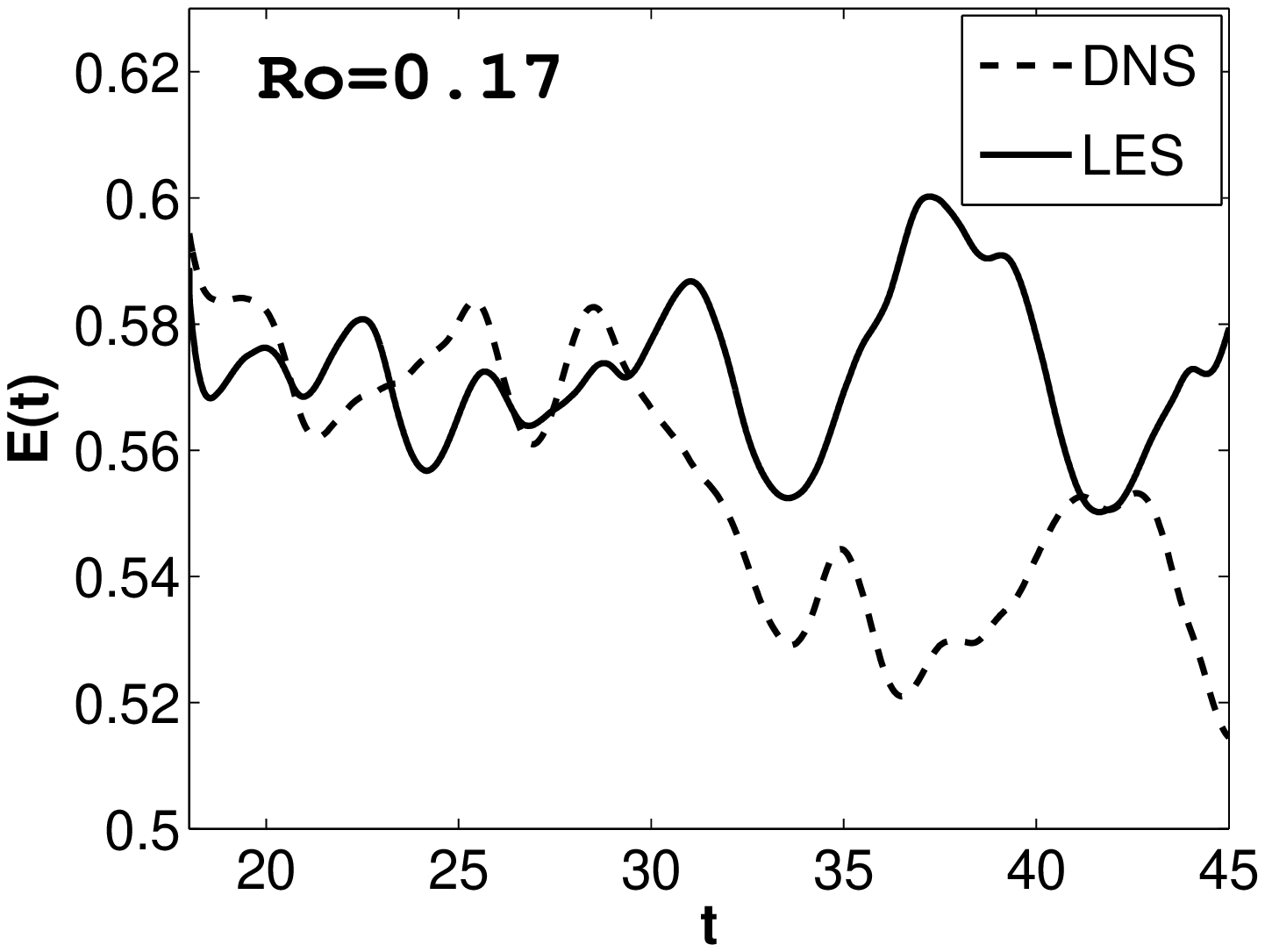}
\includegraphics[width=\linewidth,height=4cm]{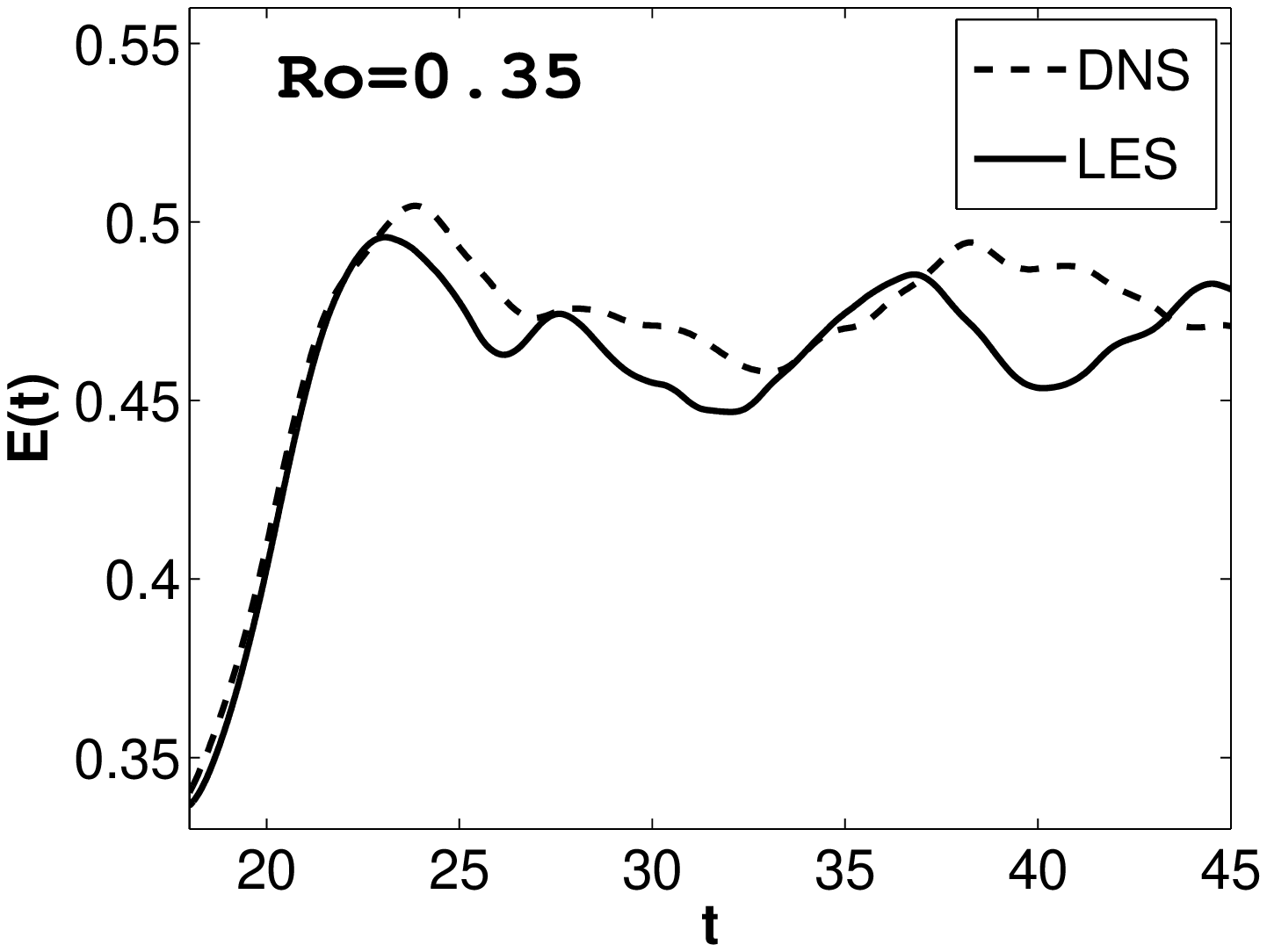}
\caption{Temporal evolution of the total energy for runs {\bf Id} (DNS: $256^3$) and {\bf IL}
(LES: $64^3$) at $Ro=0.03$ (top),
runs {\bf IId} (DNS: $256^3$) and {\bf IIL} (LES: $64^3$) at $Ro=0.17$ (middle),and runs {\bf IIId} (DNS: $256^3$) and
{\bf IIIL} (LES: $64^3$) at $Ro=0.35$ (bottom). DNS runs with dash line, LES runs with solid line. Note the change of
values on both axes for the low Rossby runs (top) because of the
delay in the development of the turbulent phase, when the LES is started, and the ensuing accumulation of energy due to the inverse cascade now taking place at that low Rossby number.
}
\label{compa_energy} \end{figure}

The first test of the model is to examine the temporal evolution of the flow. This is displayed in Fig. \ref{compa_energy}
for the three Rossby numbers analyzed. The overall behaviors of the DNS and of the LES are similar in amplitude and in time scales.
At intermediate Rossby numbers ($Ro=0.17$), the precise evolution of the DNS is not followed although the energy obtained with the LES
remains close to the DNS one. For the simulation at $Ro=0.03$, an inverse cascade develops after $t\sim120$
leading to a strong increase of the total energy. Although the LES model does not take wave interactions explicitly into account, it allows
to reproduce this transfer of energy from the small scales to the very large ones with good accuracy;
indeed, a scaling argument shows that in the small scales, the eddy turn-over time is shorter than the time associated with
waves and nonlinearities prevail.
The LES is taking into account the interactions with the waves in an implicit way by changing the EDQNM time scale
dynamically with the slope of the energy spectrum at large scales; this could be interpreted as ``reversed'' Kraichnan-like phenomenology.
Note that the run at intermediate Rossby number has higher values of the energy because the forcing amplitude
is larger than for the run at $Ro=0.35$.

\begin{figure}
\includegraphics[width=\linewidth,height=4cm]{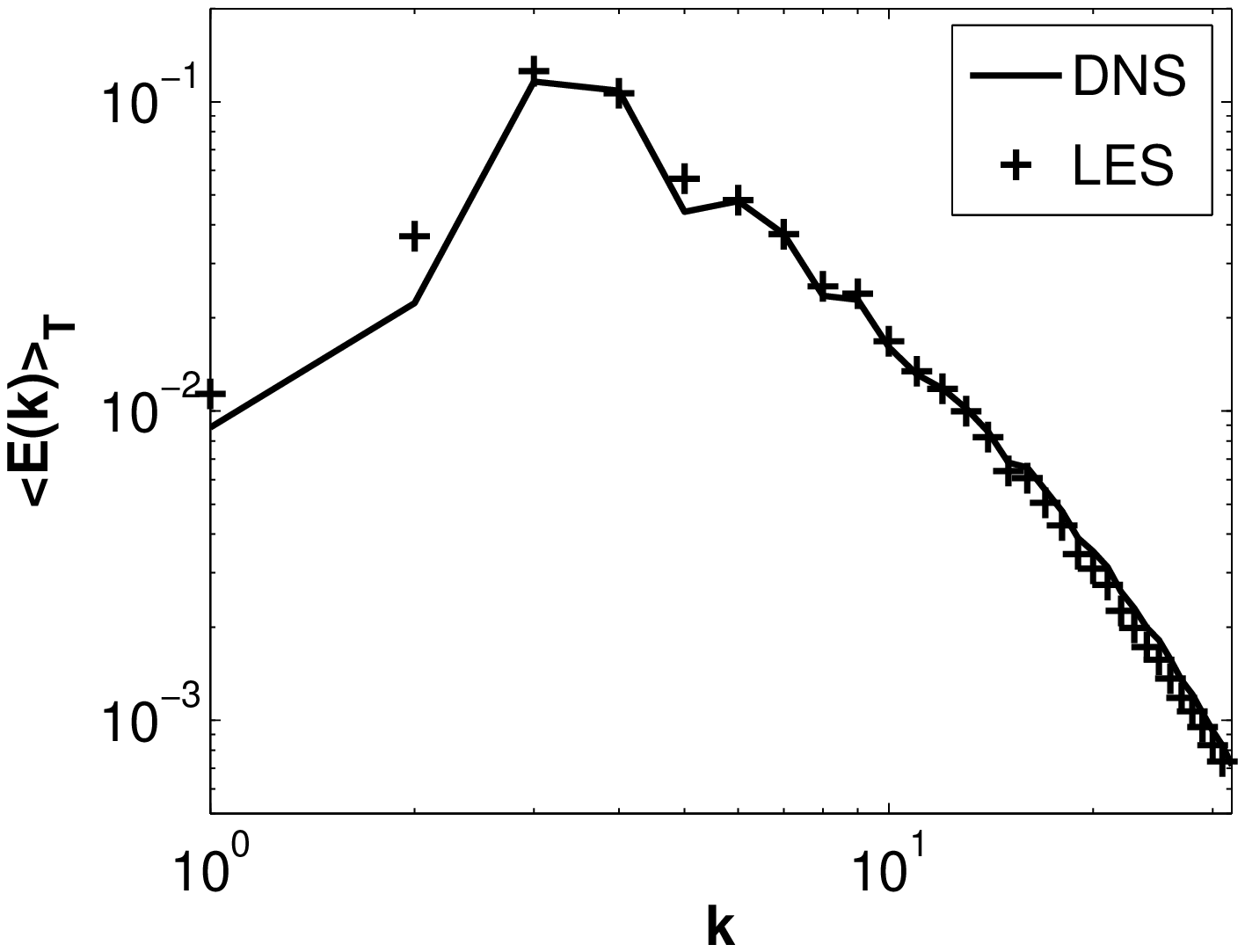}
\includegraphics[width=\linewidth,height=4cm]{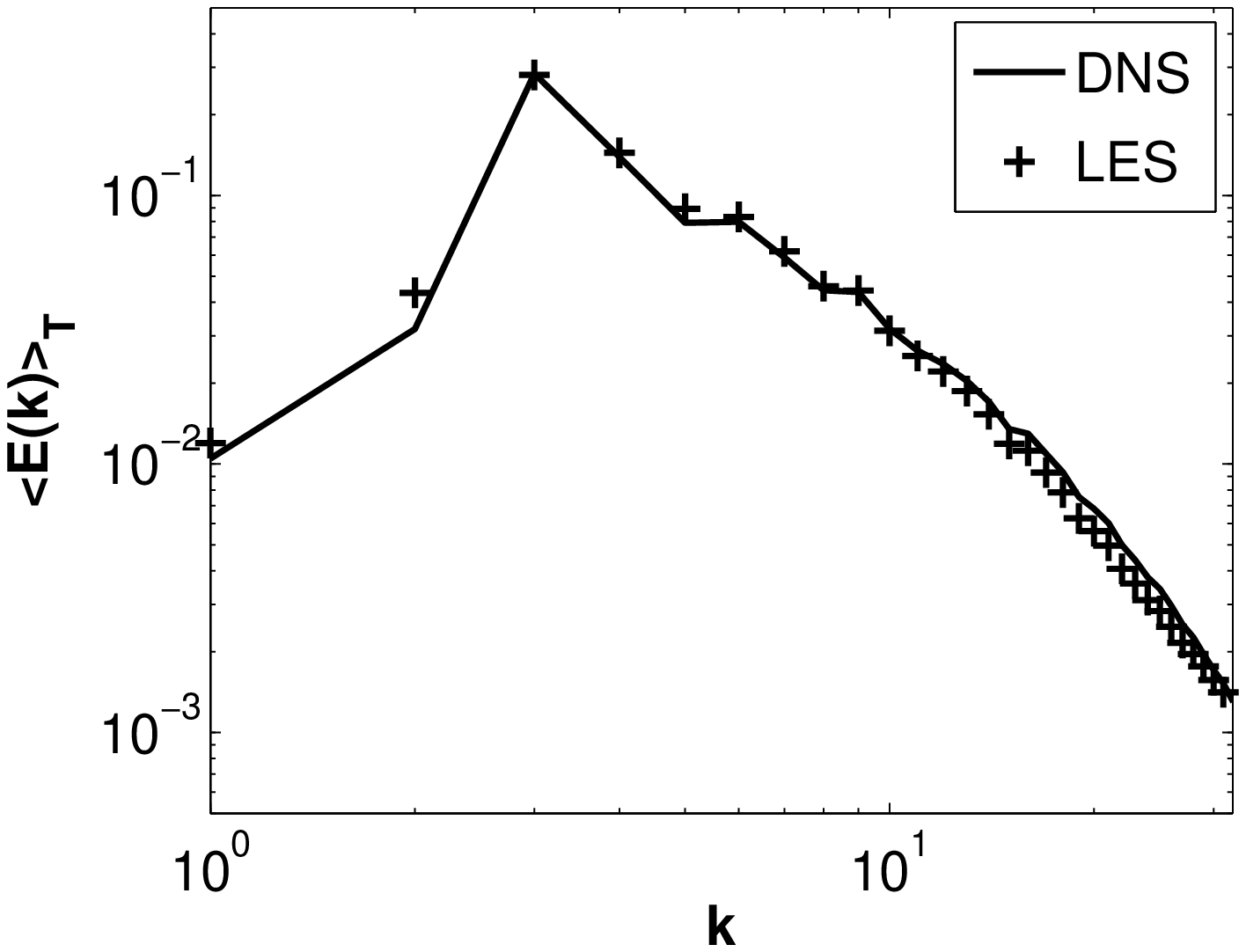}
\caption{Time averaged energy spectra for runs {\bf IId} (DNS: $256^3$) and
{\bf IIL} (LES: $64^3$) at $Ro=0.17$ (top), and runs {\bf IIId} (DNS: $256^3$) and {\bf IIIL} (LES: $64^3$)
at $Ro=0.35$ (bottom). Time averaging is performed from $t=25$ to $t=45$ for both simulations.
Note the good agreement except possibly near $k=2$ corresponding to the forcing scale, indicative of a lack of adjustment of the LES at that scale, in particular for the perpendicular spectra, see Fig. \ref{par_perp_spec}.
}
\label{compa_spectre} \end{figure}

When looking at the time-averaged isotropic energy spectra (see Fig. \ref{compa_spectre}) for the two flows at the largest
Rossby numbers, one can see that a good agreement is obtained. This figure also allows us to better understand the difference in the temporal
evolution of the energy computed from the DNS and the LES data
at $Ro=0.17$ (see Fig. \ref{compa_energy}). Indeed, although the model gives a good estimation of
the DNS spectra at small scales, at very large scale (and particularly at $k=2$)
non-negligible differences appear with the DNS, differences to which the total energy is sensitive. Note that a smaller difference
between LES and DNS spectra can be observed at $k=2$ for the run at the higher Rossby number, $Ro=0.35$.
Otherwise, the spectrum is well approximated by the LES at all the other
scales.

\begin{figure}
\begin{minipage}[t]{.49\linewidth} \begin{center}
       \includegraphics[width=4.5cm]{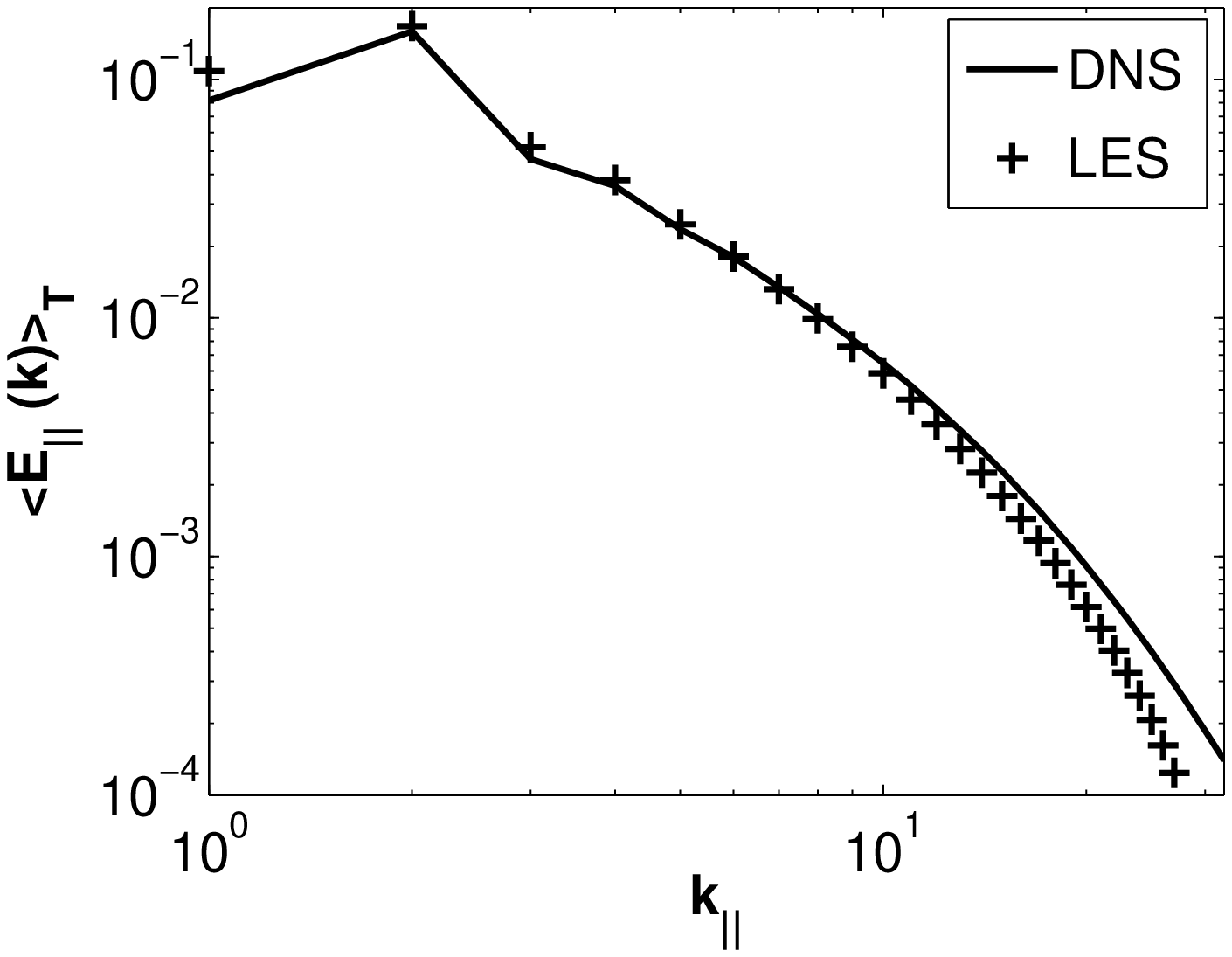}
       \includegraphics[width=4.5cm]{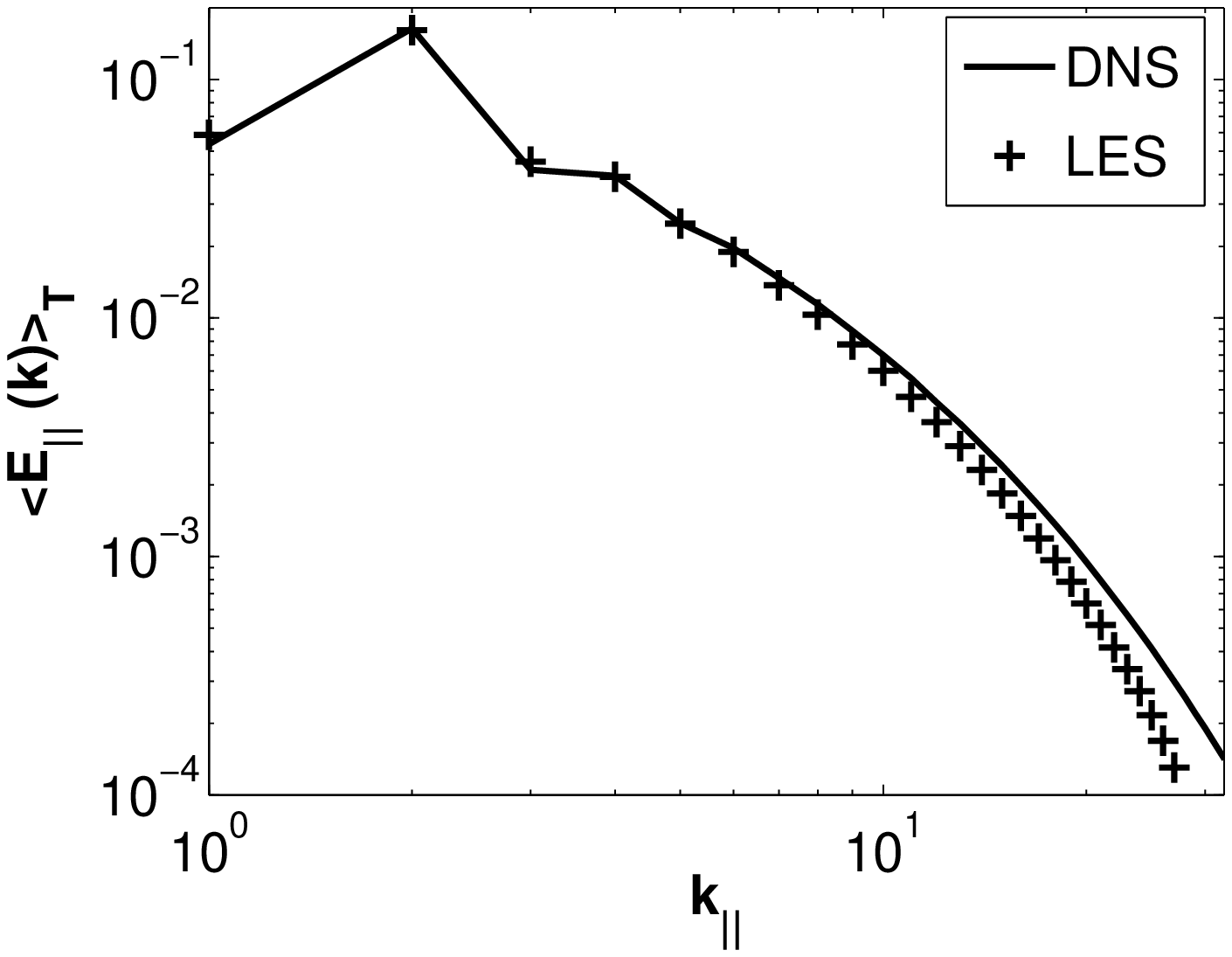}
    \end{center} \end{minipage}
\hfill
\begin{minipage}[t]{.49\linewidth} \begin{center}
       \includegraphics[width=4.5cm]{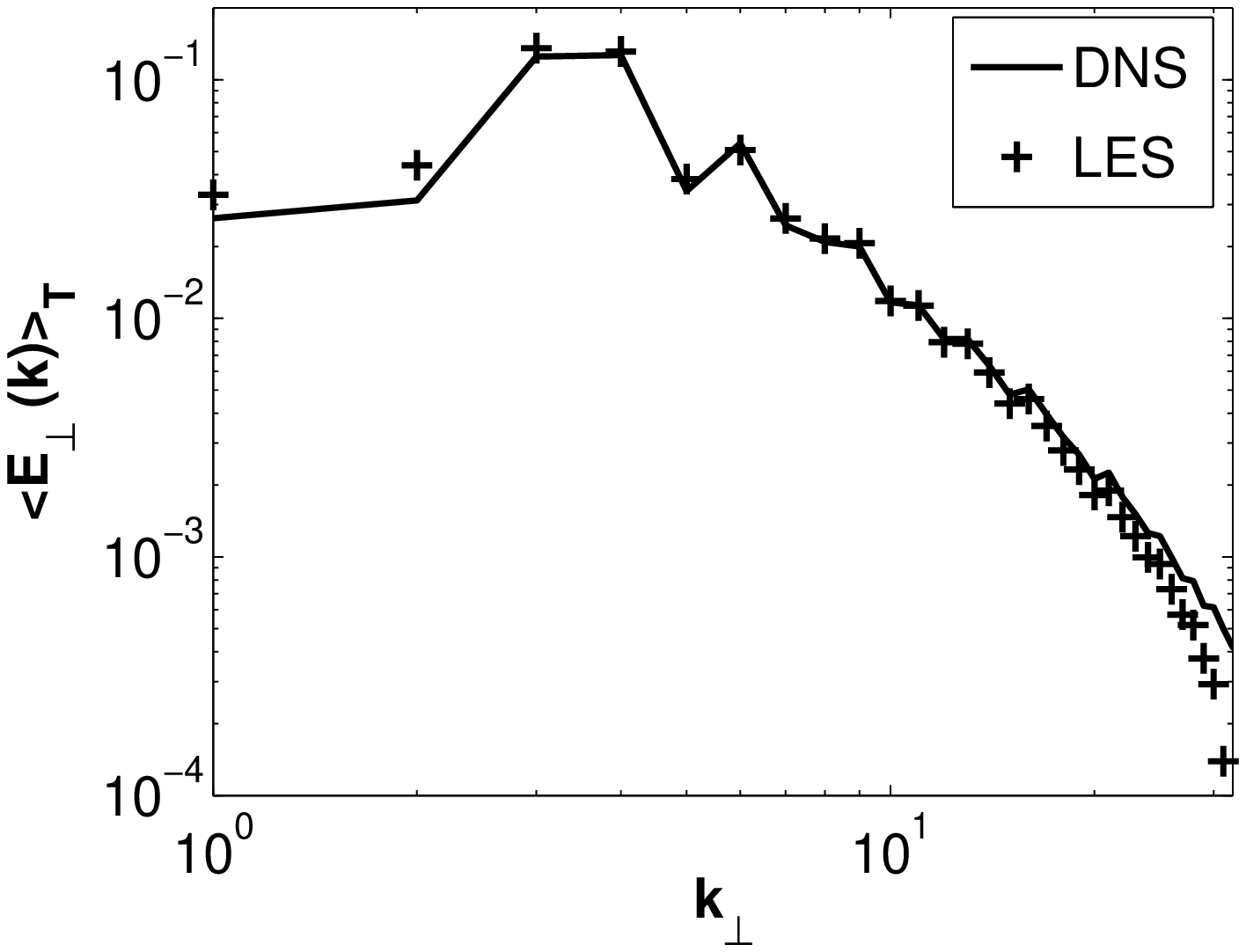}
       \includegraphics[width=4.5cm]{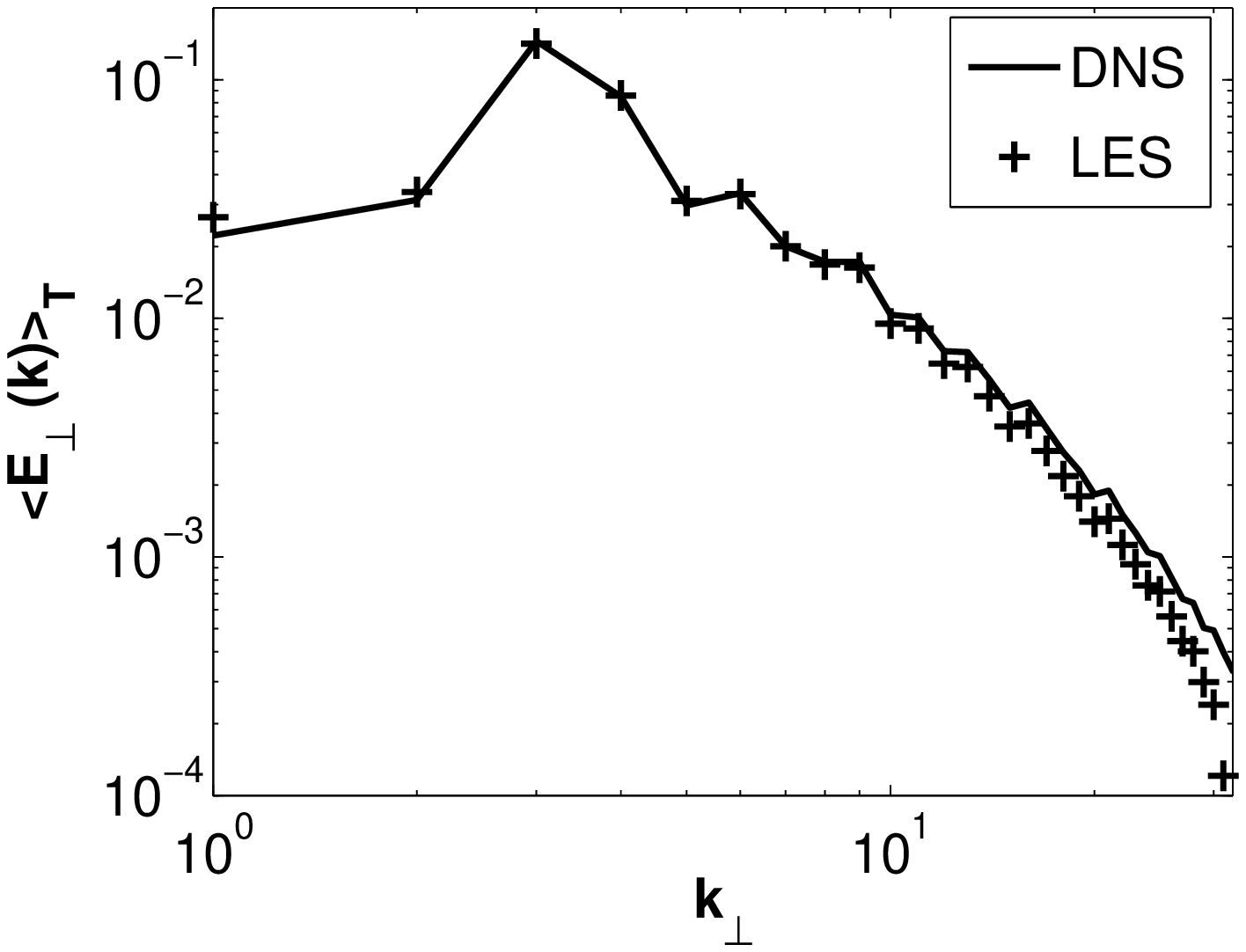}
    \end{center} \end{minipage}
\caption{Time averaged parallel (left) and perpendicular (right)
energy spectra, for runs {\bf IId} (DNS: $256^3$) and {\bf IIL} (LES: $64^3$)
at $Ro=0.17$ (top), and runs {\bf IIId} (DNS: $256^3$) and {\bf IIIL} (LES: $64^3$) at $Ro=0.35$ (bottom).
}
       \label{par_perp_spec} \end{figure}

Similarly, when decomposing the energy spectra into their perpendicular
and parallel components, a good agreement is reached at large scales, except again
at $k=2$ for the perpendicular spectrum of the flow at $Ro=0.17$
(see Fig. \ref{par_perp_spec}). On the contrary, at small scales, the model
seems to underestimate the spectra obtained by the DNS. This behavior is in
fact due to the difference in resolution between the DNS and the LES:
as $k_\perp$ and $k_\parallel$ increase, the difference between
the amount of modes taken into account in the evolution of these spectra
for the DNS and for the LES increases as well.
%\NOTE{je ne comprends pas, a un meme |K| shell, il y a a priori le meme nombre de modes ${\bf k}$, non?; d autre part, je ne comprends pas pourquoi la difference sur les spectres perp et parallel parait plus grande que sur le spectre iso}
Note that the $k_{||}$ shells have the same number of modes independently of the value of $k_{||}$
(they are planes), while the number of modes in the $k_\perp$ shells grow as $k_\perp$ (they are cylinders), and this number grows as $k^{D-1}$ in dimension $D$ for isotropic (spherical) shells. 
We have checked that,when making the comparison between the LES and the reduced DNS for instantaneous spectra, the discrepancy observed at high wavenumber disappears.

\begin{figure}
\includegraphics[width=\linewidth,height=4cm]{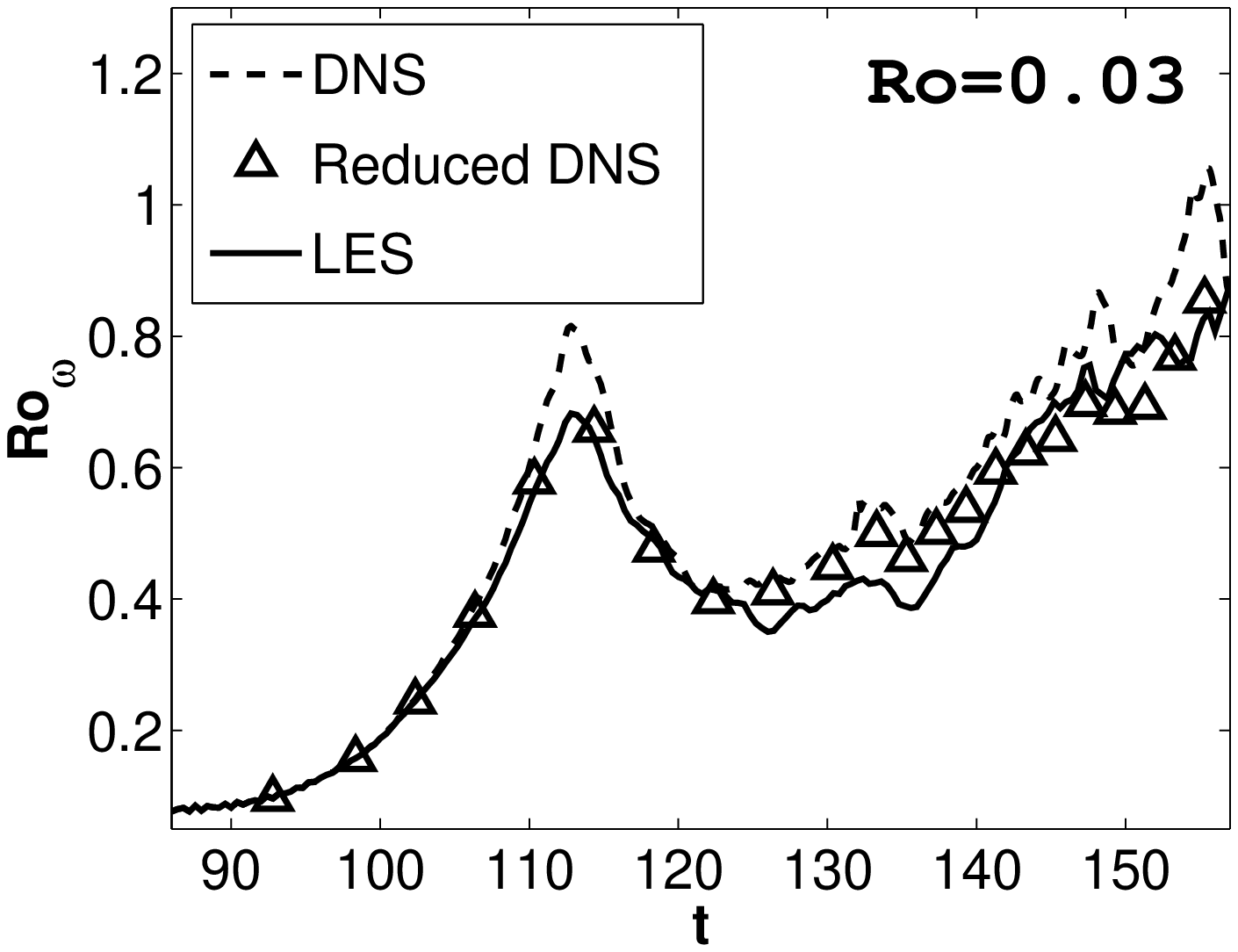}
\includegraphics[width=\linewidth,height=4cm]{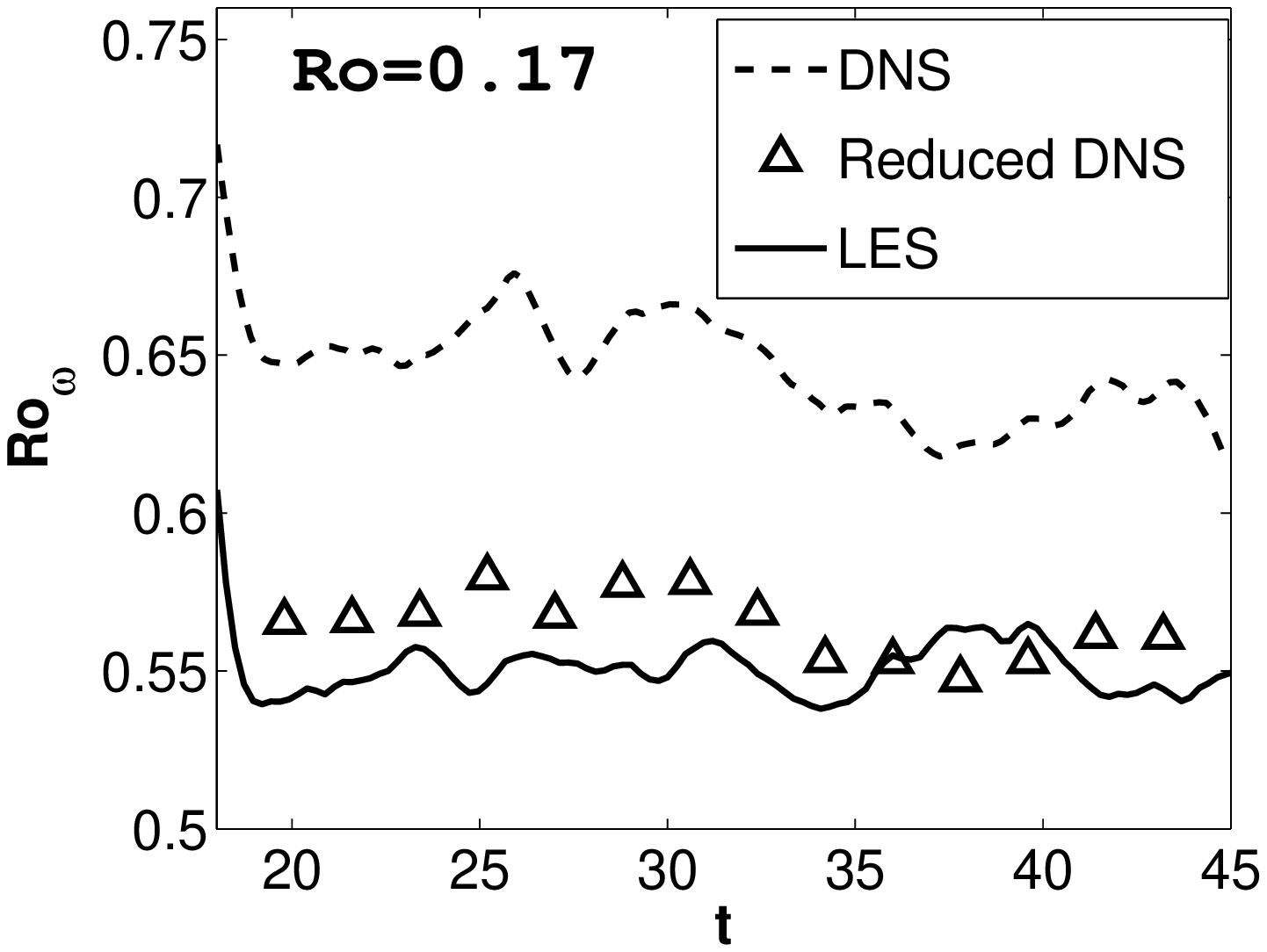}
\includegraphics[width=\linewidth,height=4cm]{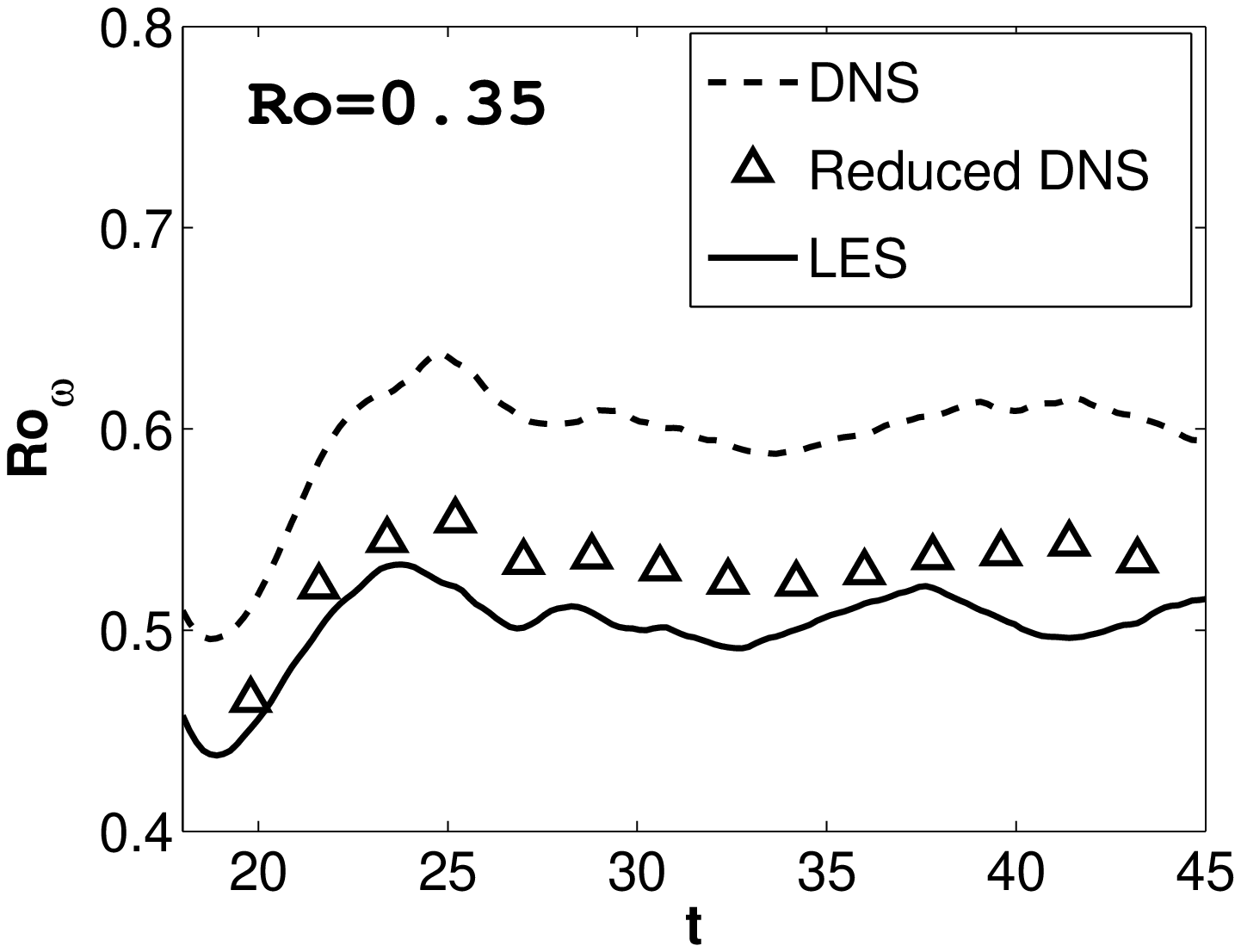}
\caption{Temporal evolution of the micro Rossby number $Ro_{\omega}$ in flows with $Ro=0.03$ (top), $Ro=0.17$ (middle) and $Ro=0.35$ (bottom),
for DNS on $256^3$ grid points (dash line), filtered data of the DNS to a $64^3$ grid (triangles) and LES (solid line) on a $64^3$ grid.
Note again the different scale on the axes, in particular for the lowest Rossby number in which case $Ro_{\omega}$ approaches unity as the inverse cascade develops and energy and turbulence intensity grow.
}
\label{compa_micro_rossby}
\end{figure}

As mentioned earlier, the micro-Rossby number measures how strong the imposed rotation is in the flow at the Taylor microscale,
when compared to the r.m.s. vorticity developed by the turbulence. Its time evolution is shown in Fig. \ref{compa_micro_rossby}
for all runs. % at $Ro=0.03$. %two large-scale Rossby numbers.
Because the micro-Rossby number emphasizes small scales that are not all present in an LES, $Ro_{\omega}$ is also computed in the reduced-DNS.
% where the small-scales have been truncated at the maximum wavenumber $N/3$ of the LES run, i.e. corresponding to a resolution of $N^3=64^3$ points.
We observe a good agreement between the truncated DNS and the LES, although the model slightly underestimates
$Ro_{\omega}$ for the two simulations at larger Rossby number.
%This behavior can probably be explained by an enstrophy transfer from the resolved to the subgrid scales which is slightly too strong. \NOTE{I don't like this last sentence, enstrophy is not an invariant and as a result its transfer is not well defined (there is production of enstrophy). I would say something like ``
This behavior can be explained by enstrophy production in the LES, and the backscattering of energy
from sub-grid scales to resolved scale associated to the eddy noise, which is perhaps not strong enough.

In Table \ref{table2} we give the values of the characteristic parallel and perpendicular
integral length scales (respectively $L_\parallel$ and $L_\perp$) defined as:
\begin{eqnarray}
L_\parallel &=& \frac{\int_1^{k_{max}} E(k_\parallel)k_\parallel^{-1} dk_\parallel}{\int_1^{k_{max}} E(k_\parallel)dk_\parallel} \ ,\label{lpara} \\
L_\perp &=& \frac{\int_1^{k_{max}} E(k_\perp)k_\perp^{-1} dk_\perp}{\int_1^{k_{max}} E(k_\perp)dk_\perp} \ ,\label{lperp}
\end{eqnarray}
and computed at the final simulation time of each flow (note that the $k_{\parallel}=0$ mode is not included in the definition).
%for each flow at late times (note that the $k_{\parallel}=0$ mode is not included in the definition).
Even if the values obtained by the LES data do not exactly correspond to the DNS values, they remain close; their
difference can be explained by the same argument evoked before on the slight
discrepancy between LES and DNS parallel and perpendicular energy spectra.
Note that the perpendicular length scale is significantly larger for the lowest Rossby number,
but the parallel length scales are comparable in all three runs. This is linked to the fact that the inverse
cascade of energy which takes place at low Rossby number is dominated by quasi-two-dimensional modes; the parallel
spectrum does not undergo an inverse cascade,
although energy does pile up at $k_{\parallel}=0$ mainly through resonant coupling of waves.

\begin{table}
\caption{\label{table2} Characteristic integral length scales $L_\perp$
and $L_\parallel$ measured at different times $t_m$ for the three
different Rossby numbers studied in this paper. Note that, at the lowest Rossby number
($Ro=0.03$, runs I), the perpendicular integral length scale is significantly larger than for more moderate rotation, because of the inverse cascade.
}
\begin{ruledtabular}
\begin{tabular}{cccccc}
 & & $t_m$ & $L_\perp$ & $L_\parallel$ \\
\hline
{\bf Id} & DNS & $157$ & $5.73$ & $2.99$ \\
{\bf IL}& LES & $157$ & $5.63$ & $2.95$ \\
\hline
{\bf IId} & DNS & $45$ & $1.71$ & $2.95$ \\
{\bf IIL} & LES & $45$ & $1.74$ & $3.22$ \\
\hline
{\bf IIId} & DNS & $45$ & $1.76$ & $2.69$ \\
{\bf IIIL}& LES & $45$ & $1.83$ & $2.71$ \\
\end{tabular} \end{ruledtabular} \end{table}

\subsection{Measures of anisotropy} \label{aniso}

\begin{figure}
\includegraphics[width=\linewidth,height=4.5cm]{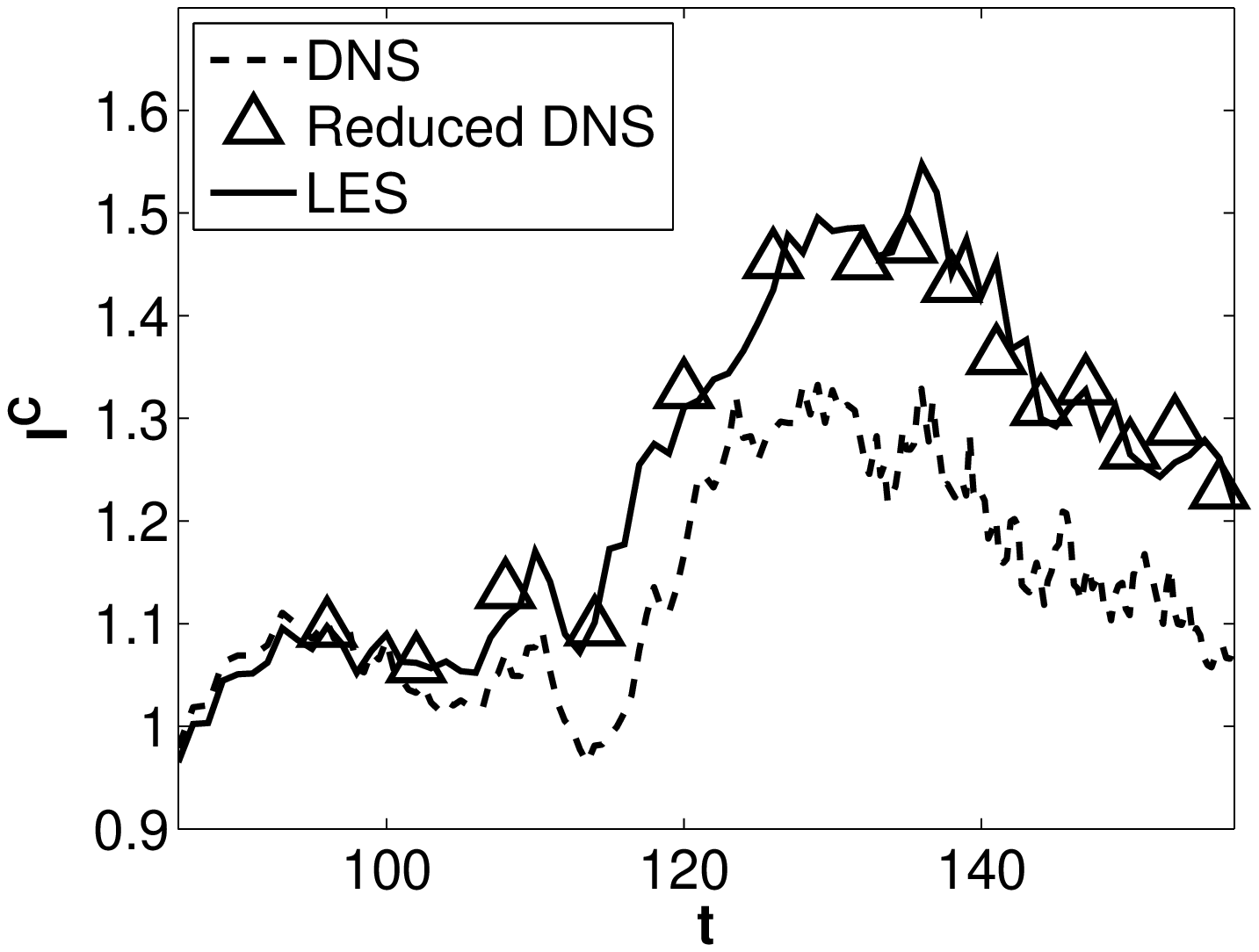}
\includegraphics[width=\linewidth,height=4.5cm]{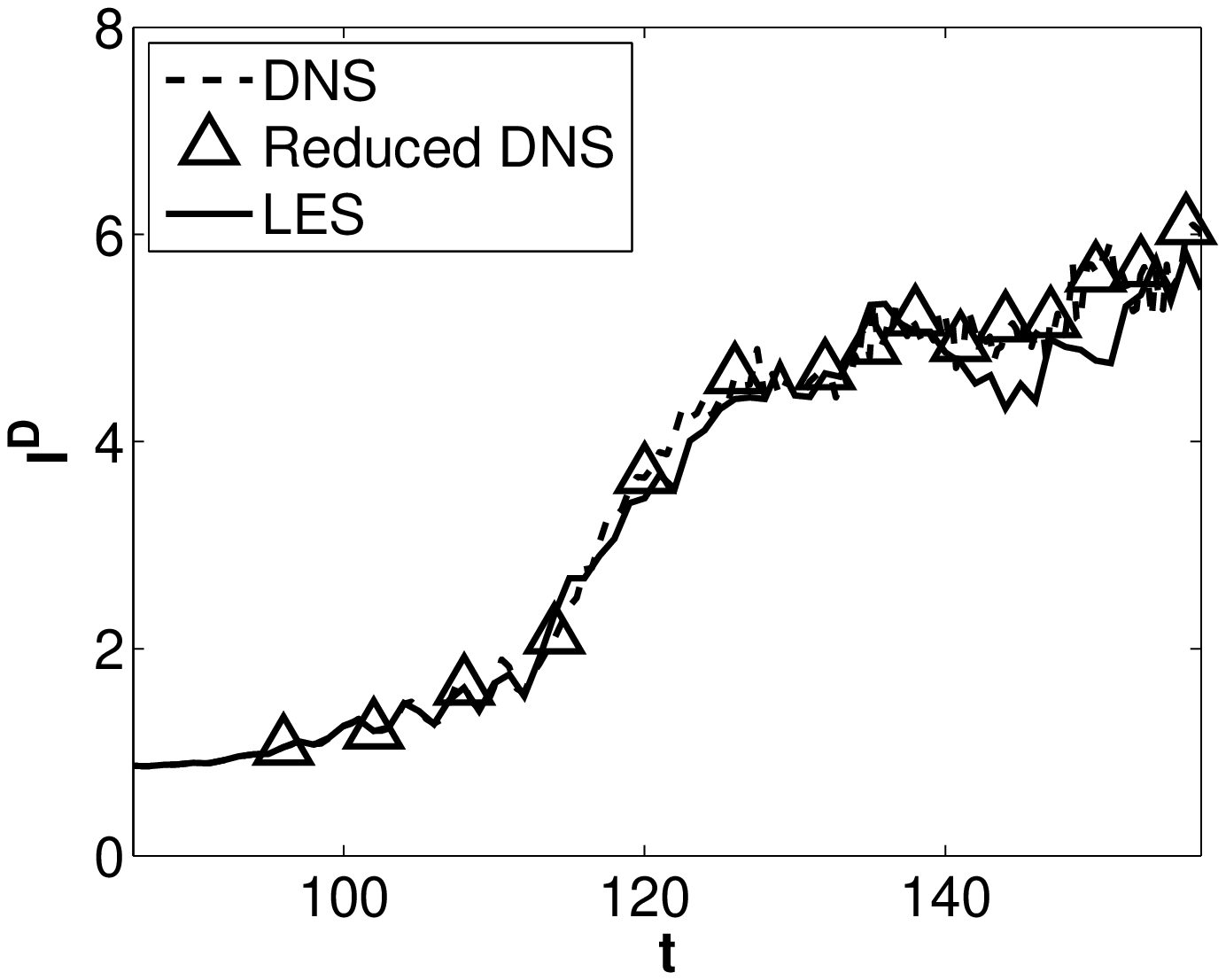}
\caption{Evolution of the spectral and directional isotropy coefficients $I^C$ and $I^D$ (see text) for runs {\bf Ir}
(reduced DNS: $64^3$) and {\bf IL} (LES: $64^3$) at low Rossby number ($Ro=0.03$).
Information about the lower rotation cases is given in Table \ref{tab3}.
Isotropy obtains when both coefficients are close to unity, and we note that the directional coefficient, related to real-space structures, indicates a stronger departure from isotropy than when measuring anisotropy in Fourier space as $I_C$ does (see Eqs. \ref{icr}, \ref{ici}), once the inverse cascade sets up and strong columnar vortices develop. Larger $I^D$ also obtain for the runs at lower Rossby numbers (see Table \ref{tab3}).
}
%\caption{Temporal evolution of the spectral and directional isotropy coefficients $I^C$ and $I^D$ (defined in the text) for runs {\bf Ir}
%(reduced DNS: $64^3$) and {\bf IL} (LES: $64^3$) at low Rossby number.}
\label{ICIR} \end{figure}

Rotating flows are known to develop anisotropies and we now turn our attention to this point.
In order to estimate the anisotropy of the different flows, we use
the coefficients $I^D$ and $I^C$ defined earlier in Eqs.
(\ref{icr}) and (\ref{ici}).
They are shown as a function of time in Fig. \ref{ICIR} for the DNS (dash line),
the reduced DNS data truncated to the LES resolution (triangles), and the LES (solid line) with $Ro=0.03$.
A very good match can be observed between the Craya
coefficient $I^C$ computed from the reduced-DNS data and the one
computed with the data from the LES model, whereas the coefficient
computed with the full DNS data evolves on a lower level than the two
other ones. This is due to the fact that the small scales
of the field (i.e. scales with $k>k_c$) are taken into account in the
spatial averaging process we perform to calculate this coefficient.
We saw in Section \ref{rot_an} that these small scales are more isotropic
with a corresponding coefficient $I^C$ near unity, so when they
are taken into account in the computation of the Craya coefficient
they lower its value. The small scales in the DNS are more isotropic, and as a result, the LES flow, which preserves a smaller amount of these scales, is globally more anisotropic and has a larger value of this coefficient.

As already observed in Fig. \ref{icr_ici}, the directional coefficient
$I^D$ is strongly dominated by the large scales of the field, such as
columnar structures appearing in the flow as a result of the inverse cascade process.
Therefore, when we compare the time history of this coefficient for
the DNS and the reduced-DNS, no noticeable difference appears. Once
again our LES model predicts very well the evolution of this coefficient, even though
 the perpendicular component of the velocity clearly dominates over the parallel one.
 %\NOTE{I removed ``indicating the development of bi-dimensional structures within the flow.'' This sentence could be added in the previous paragraph, I think the development of 2D structures is mainly measured by $I^C$ instead of $I^D$.}
%\BB{OK}
We also note that the model allows for a good estimation of both these coefficients
for the simulations at larger Rossby numbers ($Ro=0.17$ and $Ro=0.35$), as shown
in Table \ref{table3}.

\begin{table}
\caption{\label{table3} Craya and directional isotropic coefficients $I^C$ and $I^D$ for the simulations
at $Ro=0.17$ and $Ro=0.35$.
}
\begin{ruledtabular}
\begin{tabular}{cccccc}
 & & $t$ & $Ro$ & $I^C$ & $I^D$ \\
\hline
{\bf IId} & DNS & $45$ & $0.17$ & $1.05$ & $1.69$ \\
{\bf IIr}& Reduced DNS & $45$ & $0.17$ & $1.07$ & $1.69$ \\
{\bf IIL} & LES & $45$ & $0.17$ & $1.07$ & $1.71$ \\
\hline
{\bf IIId} & DNS & $45$ & $0.35$ & $1.04$ & $1.97$ \\
{\bf IIIr}& Reduced DNS& $45$ & $0.35$ & $1.04$ & $1.97$ \\
{\bf IIIL}& LES & $45$ & $0.35$ & $1.04$ & $2.01$
\label{tab3} \end{tabular} \end{ruledtabular} \end{table}

\begin{figure}
\includegraphics[width=\linewidth,height=4.5cm]{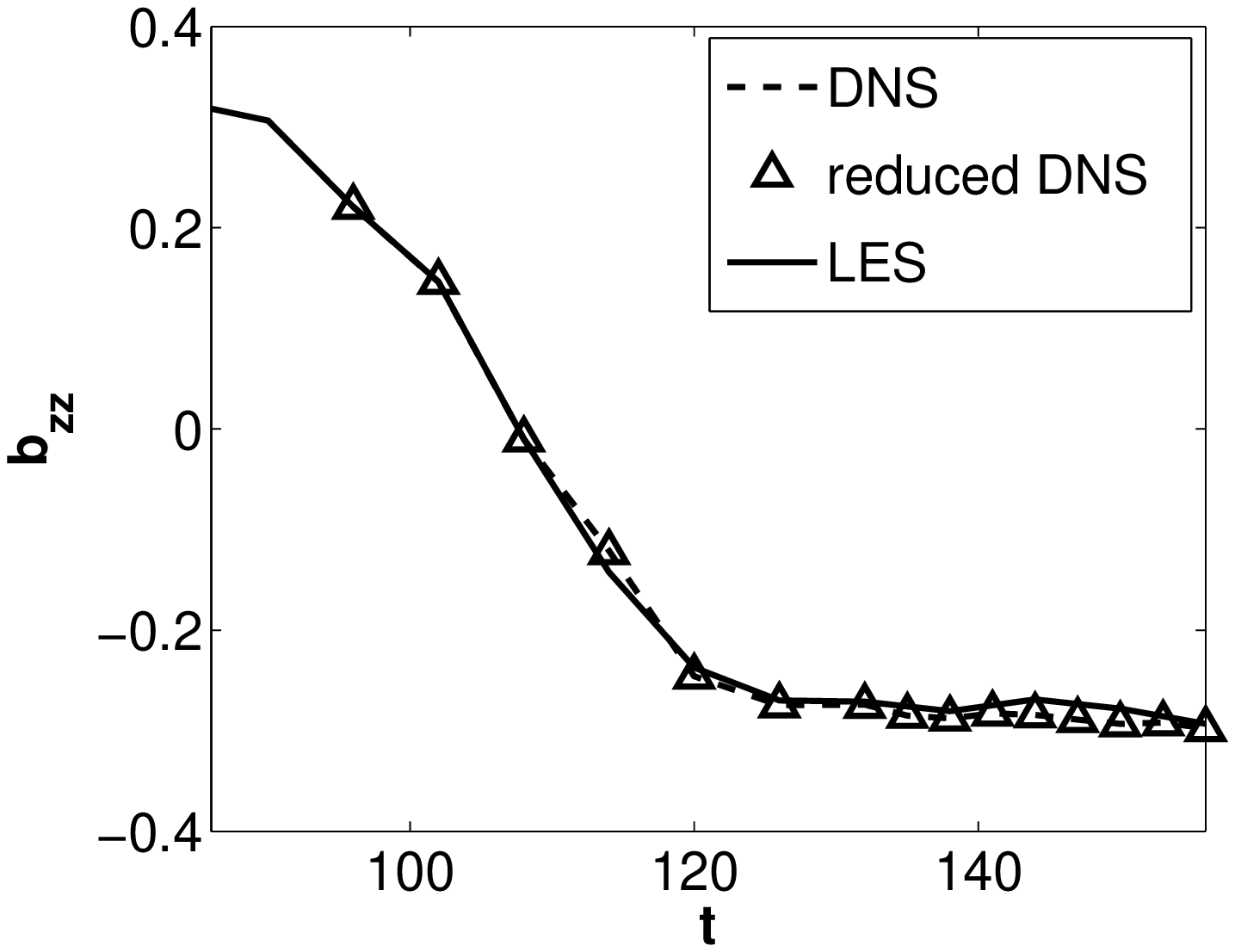}
\includegraphics[width=\linewidth,height=4.5cm]{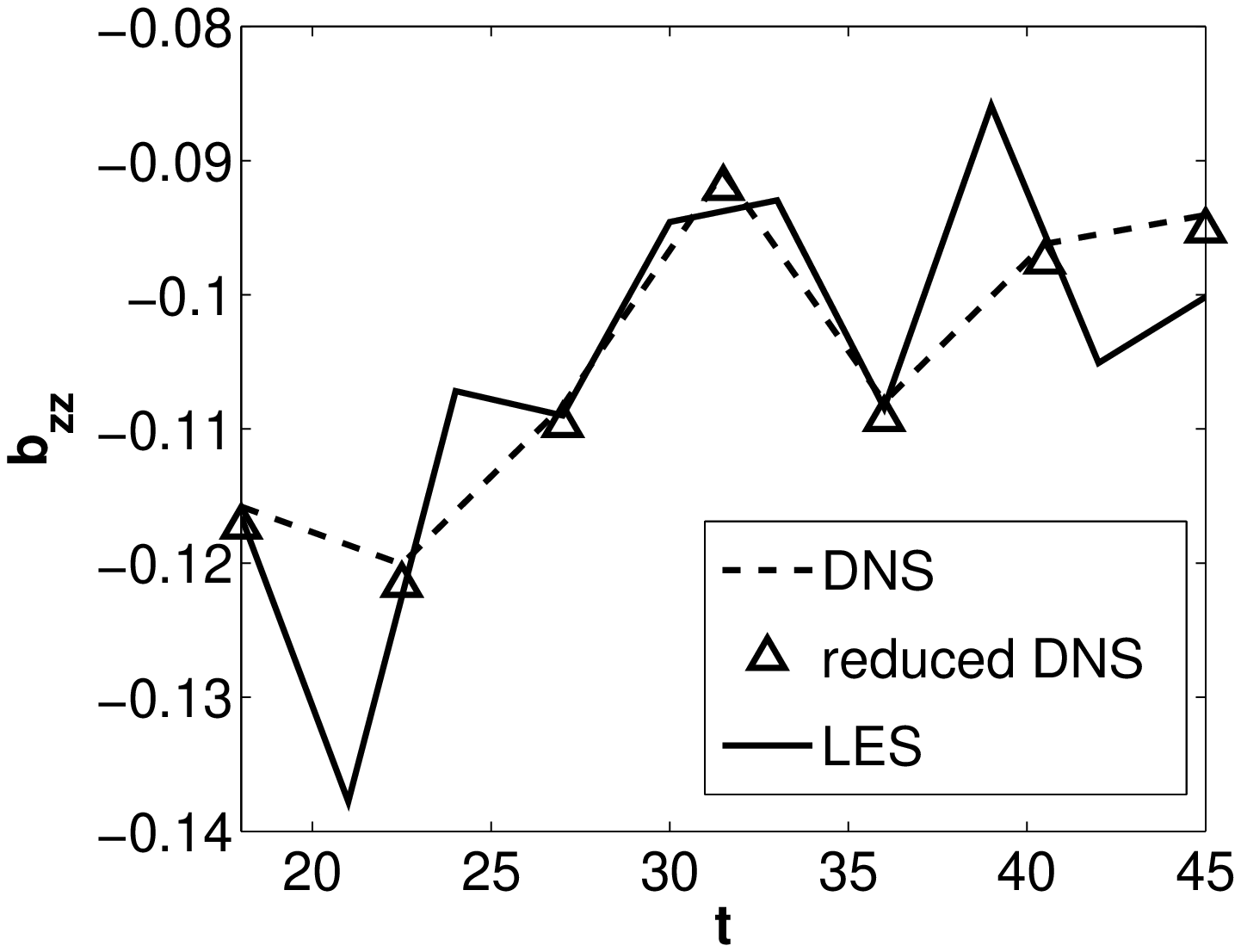}
\caption{Temporal evolution of the $b_{zz}$ component of the polarization anisotropy tensor (see Eq. \ref{BZZ}) for runs
with $Ro=0.03$ (top) and $Ro=0.35$ (bottom), in the former case at late times once the pile-up of energy at large scales has begun. The DNS are plotted with a dash line, the reduced DNS, truncated to the resolution of the LES, are given with triangles and the solid line corresponds to the LES. Note again the tendency to a two-dimensional state at late times, with $b_{zz}\rightarrow -1/3$, for the low Rossby number runs.}
\label{b33} \end{figure}

In our investigation of anisotropy of rotating flows, we finally study the behavior
of the $b_{ij}$ anisotropy tensor defined below (see e.g. \cite{sagaut} for reference); it is linked to the so-called ``polarization'' anisotropy introduced in \cite{cambon_jacquin} and as also discussed in \cite{morinishi} (see also \cite{yang}).
This tensor, which is based on the Reynolds stress tensor $R_{ij} = \langle v_i(\bx) v_j(\bx)\rangle$,
is defined as:
\begin{equation}
b_{ij} = \frac{R_{ij}}{R_{ll}} - \frac{\delta_{ij}}{3},
\label{BZZ}\end{equation}
with summation upon the subscript $l$.
In Fig. \ref{b33} we represent the temporal evolution of the
$b_{zz}$ component of the anisotropy tensor for runs {\bf I} and {\bf III}, at
respectively $Ro=0.03$ and $Ro=0.35$.
We first notice that the LES model predicts well the evolution
of this coefficient for both simulations. Secondly, the development
of a preferred direction in the flow at $Ro=0.03$ (already observed in Fig. \ref{ICIR} through the increase of
the directional coefficient $I^D$ in the inverse cascade), is also visible in this figure.
Indeed, $b_{zz}$ tends to $-1/3$ as time increases, since $v_z(\bx)$ becomes negligible  when compared to the horizontal components $v_x(\bx)$ and $v_y(\bx)$.

\subsection{Statistical analysis}\label{stat}

\begin{figure}[h]
\includegraphics[width=\linewidth,height=4cm]{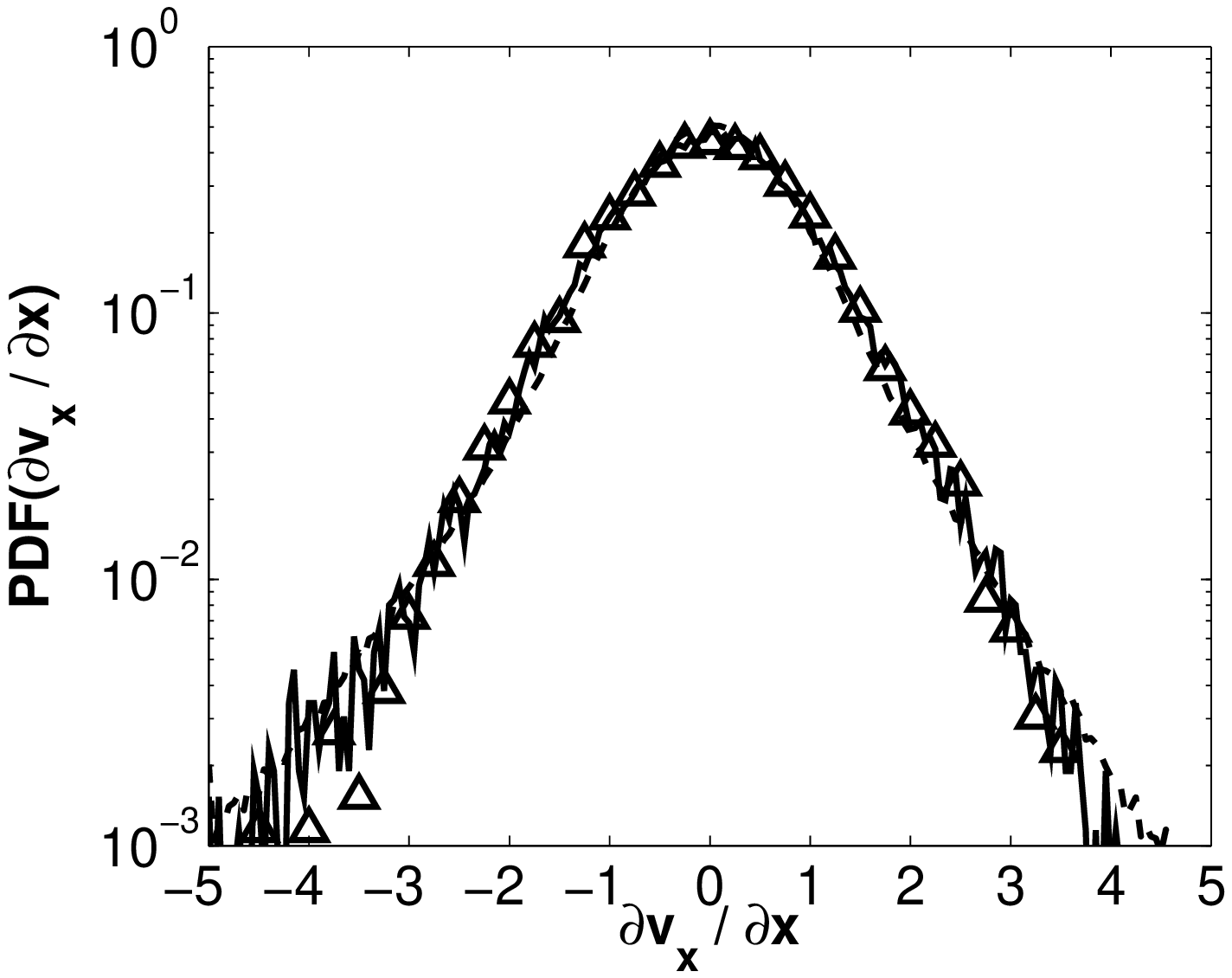}
\includegraphics[width=\linewidth,height=4cm]{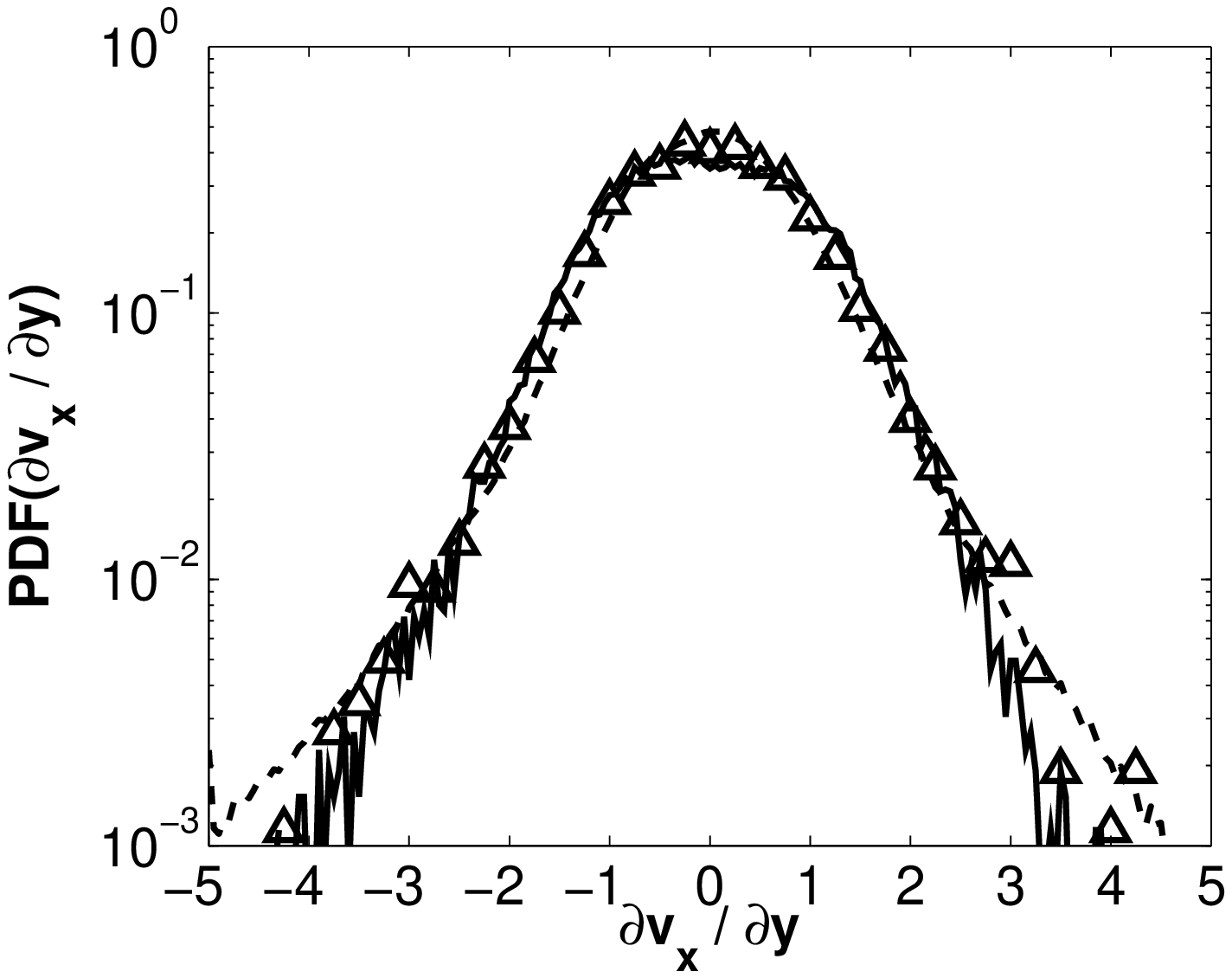}
\includegraphics[width=\linewidth,height=4cm]{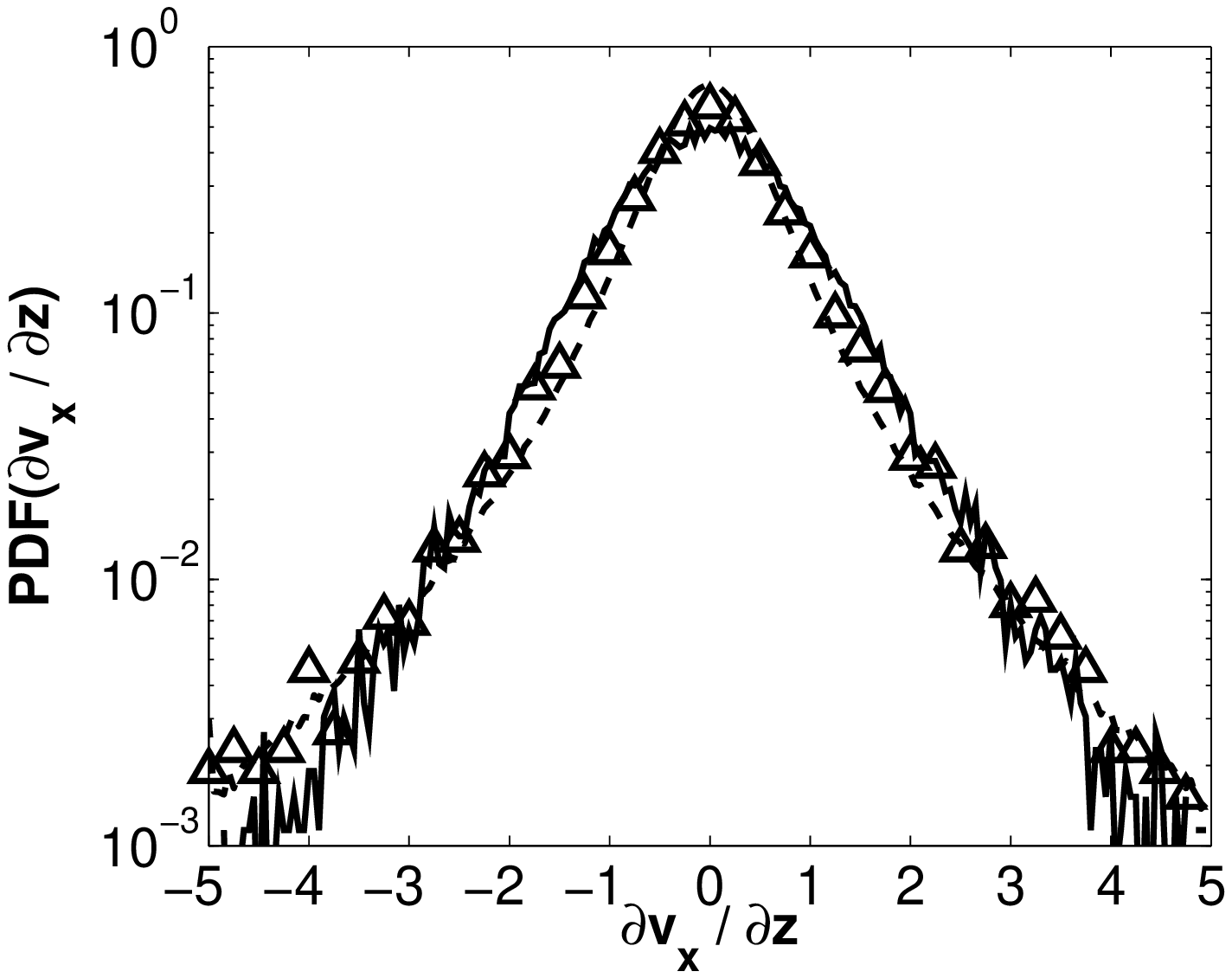}
\caption{Probability density function of the velocity derivatives
${\partial v_x}/{\partial x}$ (top),
${\partial v_x}/{\partial y}$ (middle),
and ${\partial v_x}/{\partial z}$ (bottom), for runs {\bf Id}
(DNS: $256^3$), {\bf Ir} (reduced DNS: $64^3$), and {\bf IL} (LES: $64^3$) at $Ro=0.03$ and $t=132$.
As usual, dash line is for the full DNS flow, triangles for the reduced (truncated) DNS and solid line for the LES.
}
\label{pdf} \end{figure}

In this section, we investigate the statistics of the simulations at $Ro = 0.03$.
Instantaneous probability density functions (or PDFs) of the longitudinal
and transverse derivative of the $x$-component of the velocity field are computed and
plotted in Fig. \ref{pdf} at time $t=132$, in the inverse cascade. The PDFs computed on the full DNS data, the reduced-DNS,
and the LES, agree well for the case of the longitudinal derivative. In the case of the transverse derivative,
the DNS data differ from both the LES and the reduced-DNS data, the latter
two displaying wider wings and being almost superimposed. It is well known that the small scales of
a flow may have a strong influence on the distribution of velocity derivatives with strong velocity gradients appearing at small scale,
and that transverse derivatives show stronger tails in the {\it pdfs} (and therefore enhanced intermittency) than longitudinal derivatives.
It is not clear whether this is the effect of more sensitivity to the intermittency in the transverse increments or whether it is
the effect of small-scale anisotropy, but since the differences are stronger for the velocity derivatives taken in the direction of rotation, it may be attributed to anisotropies.
%leading to the widening of the {\it pdf}'s wings.

\begin{figure}[h]
\includegraphics[width=\linewidth,height=4cm]{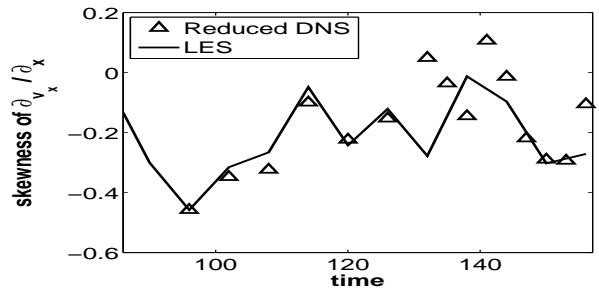}
\caption{
Temporal evolution of the skewness for the longitudinal velocity derivative ${\partial v_x}/{\partial x}$ (see Eq. \ref{SKEW}),
for runs {\bf Ir} (reduced DNS: $64^3$) and {\bf IL} (LES: $64^3$) at $Ro=0.03$.
%The skewness of a field $f(\bx,t)$ is defined as $S_3=\left<f^3\right>/\left<f^2\right>^{3/2}$.
%Temporal evolution of the skewness of the velocity derivative ${\partial v_x}/{\partial x}$,
%%${\partial v_y}/{\partial y}$ (middle),
%for runs {\bf Ir} (reduced DNS: $64^3$) and {\bf IL} (LES: $64^3$) at $Ro=0.03$.
%The skewness of a field $\bv (\bx,t)$ is defined as $S_3=\left<v^3\right>/\left<v^2\right>^{3/2}$.
}
\label{skewness} \end{figure}

In order to quantify the distributions of velocity fluctuations and the differences
between DNS and LES data, we now compute the skewness $S_k$ of the longitudinal velocity derivative, i.e. its normalized third order moment.
The skewness, which measures the departure from Gaussian statistics,
 is usually negative for the longitudinal derivatives of a turbulent flow
and oscillates around zero for the lateral ones. In Fig. \ref{skewness}
we show the time history of $S_k$.
%we show the time history of $S_3\left(\frac{\partial v_x}{\partial x} \right)$.
As for the energy, the LES model gives a correct prediction of the skewness for
$86 < t < 145$, although around $t\simeq 140$, some discrepancy can be found that could be associated
with the development of structures.
Note that this difference can be also associated with a slight discrepancy in total energy at around the same time (see Fig. \ref{compa_energy}).
% and $t > 145$, but between these two ranges the model solution diverges from the DNS one. Unfortunately we did not find any explanation to this phenomenon.

\subsection{Visualization in physical space}\label{stat}

\begin{figure}
\includegraphics[width=\linewidth]{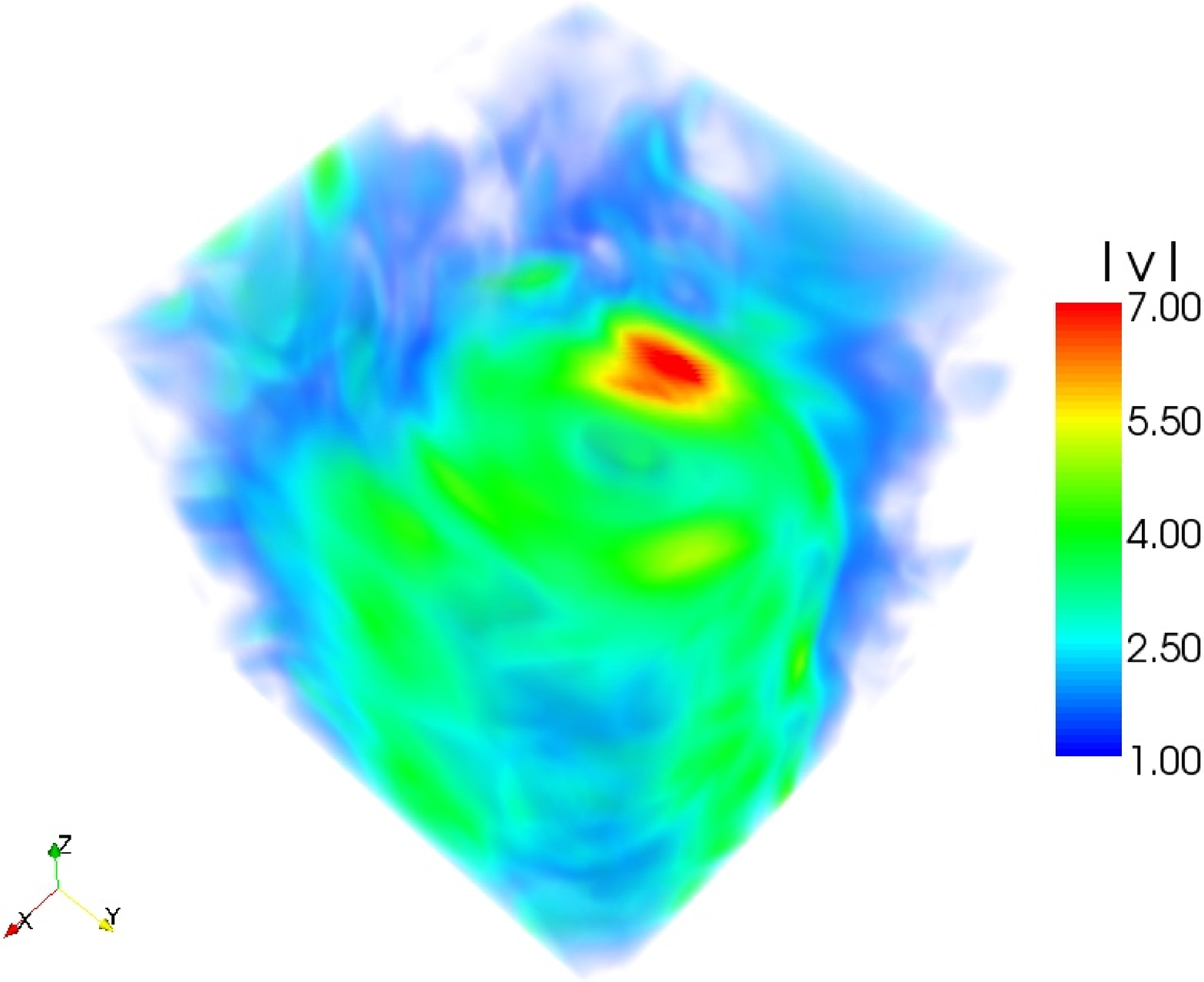}
\includegraphics[width=\linewidth]{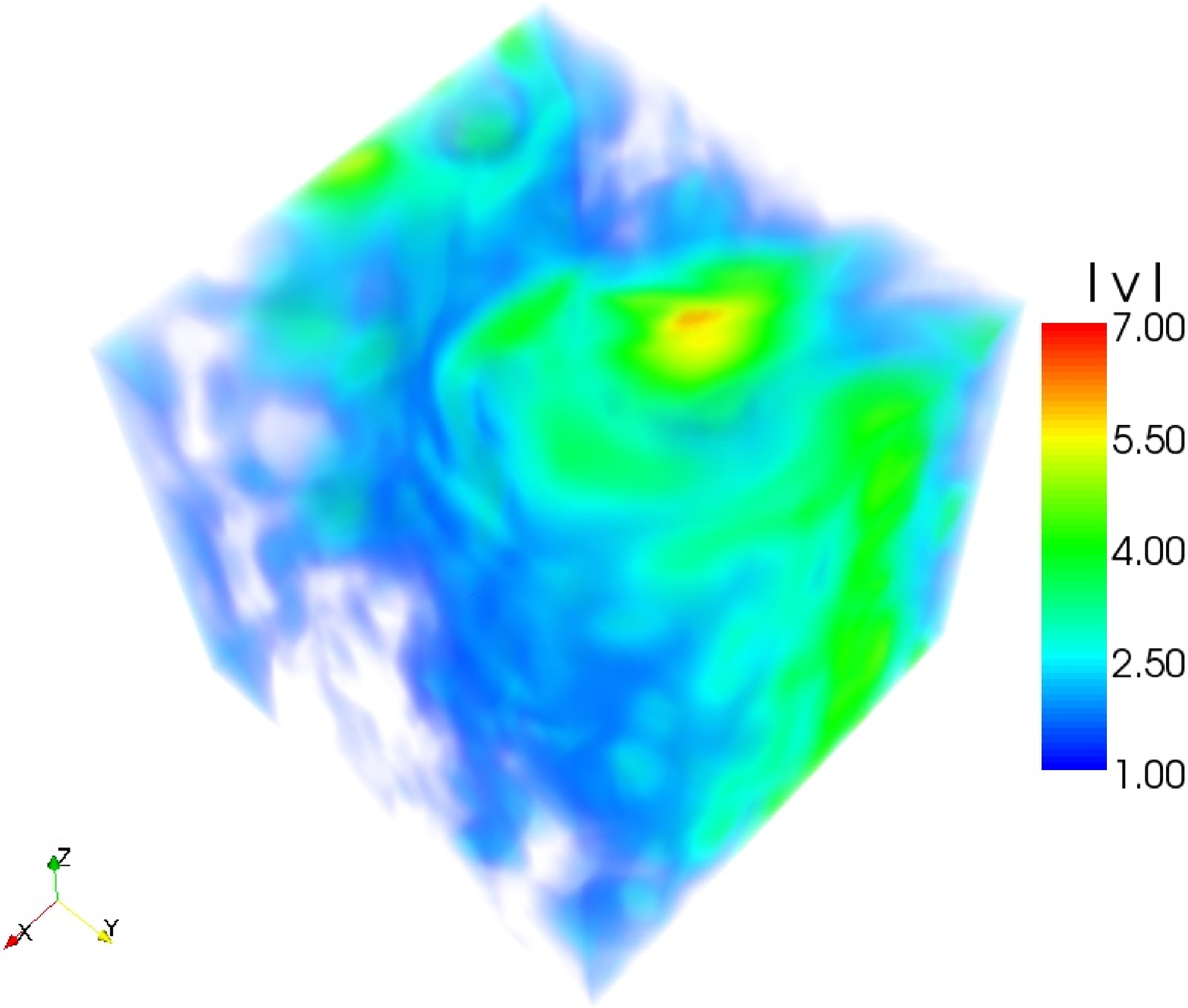}
\caption{Volume rendering of the velocity modulus for flows at $Ro=0.03$ -- runs {\bf Ir} (Reduced
DNS: $64^3$, top), and {\bf IL} (LES: $64^3$ bottom) -- at time $t=132$ for both simulations. The flow is dominated by a large eddy, but 
smaller vertical structures are visible as well.
}
\label{volume} \end{figure}

We finally present a visualization in physical space of
the velocity intensity at $t=132$ for the flow at $Ro=0.03$.
At this time of the simulation the inverse cascade already took place
and most of the flow energy was transferred to the $k_\parallel = 0$ plane.
We noted earlier that the TG flow injects no energy in the $k_\perp=1$ shell nor in the $k_{||}=0$ shell.
So all energy we see at large scale is the result of inverse cascade (in the former case) and of two-dimensionalization
(in the latter case). The evidence for the inverse cascade in this paper is given by the time evolution of the energy in Figs. 1 and 3 (see also Paper I, where fluxes are studied in detail).
The accumulation of energy in this plane leads to the formation
of columns as can be observed in Fig. \ref{volume}. Although the structures
are quasi-bidimensional, the isotropic LES model allows to reproduce them quite correctly.
The spatial position differs slightly from the structure obtained by the DNS, but its size and
intensity are well approximated. 
When examining the temporal evolution of the maximum of velocity (not shown), a good agreement occurs at all times.
Note that this is a forced run visualized after $\approx 130$ turnover times; as a result of 
the intrinsic sensitivity of turbulent flows due to their inherent unpredictability after a Lyapounov time of the order of a few turn-over times,
% sensitivity to initial conditions in a turbulent flow, 
the spatial position of the structures is not expected to be reproduced exactly by the LES.

\section{Conclusion}

We present in this paper a Large Eddy Simulation model for high Reynolds number rotating flows using a previously derived
sub-grid model \cite{LESPH} (see also \cite{JB_PHD}) based on the isotropic EDQNM two-point closure with eddy viscosity and eddy noise.
We show that, down to Rossby numbers of $0.03$, the small-scales are sufficiently isotropic for the model to perform reasonably well.
There are numerous laboratory experiments with which the comparison presented here could be extended, following the work
in \cite{Cambon97} using several EDQNM-based closures (see also \cite{godeferd03} for the stratified anisotropic case).
% and giving for decaying flows a general agreement.
The advantage of using two-point closures such as EDQNM as a model for turbulent flows in the presence of rotation is that it allows
for building scaling laws at a relatively low computational cost and with the possibility of doing analytical estimations
of nonlinear transfer (see for example \cite{Cambon97}).
The model presented in this paper is much simpler since it is built on the isotropic three-dimensional version of the EDQNM;
it is thus more limited in its scope insofar as it may not be able to explore very low Rossby numbers.
On the other hand, following the standard LES methodology with spatially resolved large scales, and turbulent coefficients to model the sub-grid fluctuations, it allows to access more detailed features of the flows such as high-order statistics
(e.g., PDFs) as well as spatial structures.
%Both methodologies should be employed to pursue the investigations of these important flows.

Also, the LES used in this paper adapts dynamically depending on the spectral index of the energy at super-filter
(resolved) scales, and the value of the turbulent transport coefficients vary as a result.
This is important in the context of rotating turbulent flows because the power law followed by the energy spectrum
in this case is not quite ascertained yet and does vary with time. Phenomenological and theoretical predictions of this index,
as well as several recent experiments, were reviewed in \cite{morize}, with experimental and numerical evidence not quite able yet
to sort out the different models or to fully describe the parameter space (e.g., as function of the rotation rate $\Omega$, the Reynolds
number, etc.). An adequate LES model that can adjust to the resolved energy spectrum can help in this matter but more development and
tests are needed. A reminiscent situation is found in magnetohydrodynamics (MHD) when coupling the fluid to a magnetic
field in the non-relativistic limit; the total energy spectrum obtained analytically from the weak turbulence
limit \cite{galtier00,galtier02} has been observed in the magnetosphere of Jupiter \cite{saur} and in DNS \cite{1536WT},
but the strong turbulence spectrum (or spectra in case there are different regions in parameter space) is a matter of debate.

Only one specific (non-helical) forcing was explored in the DNS-LES comparisons studied in this paper. Further tests are required,
considering other (non-helical) forcing functions, as well as forcing functions that introduce both energy and helicity in the flow.
In this latter case, the implementation of the LES as described here may prove insufficient and one should also consider taking into
account the spectral properties and turbulent transport coefficients that include the effect of helicity, as done in the non-rotating case in Paper II.
Such an implementation can also be of interest for non-helical flows, because even though helicity is not a positive definite quantity,
local helical fluctuations develop rapidly in a flow through alignment of vorticity and pressure gradients \cite{matthaeus}.
%This paper has concentrated on the validity of implementing an isotropic model of rotating flows at small but moderate Rossby numbers and has thus put aside the issue of helicity.
The properties of the model in the helical case in the presence of rotation will be dealt with in a forthcoming paper.
The freely-decaying case (see \cite{Cambon97,belletJFM08,moisy08} for a global perspective) needs to be examined as well and is left for future work.

%Compared to other models of rotating turbulence, we note for example that the assumption of locality (in Fourier space) of nonlinear interactions made in \cite{canuto} may not apply here, as measured in \cite{us_alex}. ...
%\ADD{may be we do not want to be only negative ... and we only speak about it in the introduction of the edqnm model ...}

As a final remark, we want to stress the importance of developing adequate modeling of rotating (and stratified) flows,
as encountered for example in the Earth atmosphere. It was shown recently \cite{rotunno} that the maximum intensity of a hurricane
depends crucially on the (assumed) horizontal mixing length; this implies that an adequate treatment of the turbulence is essential
in predicting various properties of hurricanes such as its intensity or landfall localization.
A run with resolution down to 62 meters shows strong local winds that were unresolved in previous studies \cite{rotunno2}.
If the work presented here (as well as most of its predecessors) is far from reality for hurricane dynamical modeling
(because of its lack of proper boundary conditions, of stratification, of moisture, ...), it nevertheless represents a first step
towards the goal of a better understanding of geophysical flows, the issue here being that sufficiently high Reynolds number,
i.e. sufficient multi-scale interactions and two-way coupling between the small scales and the large scales
in turbulent fluids supporting inertial (and/or gravity) waves, is a desired ingredient for testing LES approaches to geophysical turbulence.

\begin{acknowledgments}
Computer time was provided by NCAR which is sponsored by NSF.
PDM is a member of the Carrera del Investigador Cient\'{\i}fico of CONICET.
\end{acknowledgments}

\appendix
\section{Closure expressions of transfer terms}

For completeness, we recall here the expression of the EDQNM closure
equation for the kinetic energy spectrum $E(k,t)$ without helicity (note that the Coriolis term vanishes in the energy equation).
%, it only appears in the transfer term (of the EDQNM equations).
\begin{equation}
(\partial_t + 2\nu k^2 )E(k,t) = \widehat{T}(k,t)
\end{equation}
where the nonlinear transfer terms $\widehat{T}(k,t)$ is expressed as:

\begin{equation}
\widehat T(k,t) =
\iint_{\Delta}\theta_{_{kpq}}(t)S_E(k,p,q,t)dpdq \ .\label{teedqnm}
\end{equation}
Here $\Delta$ is the integration domain with $p$ and $q$ such that (${\bf k}, {\bf p}, {\bf q}$) form a triangle, and $\theta_{_{kpq}}(t)$ is the relaxation time of the triple velocity correlations.
As usual \cite{lesieur_book}, $\theta_{_{kpq}}(t)$ is defined as :
\begin{equation}
\theta_{_{kpq}}(t)=\frac{1-e^{-(\mu_k+\mu_q+\mu_p)t}}{\mu_k+\mu_q+\mu_p} \ ,
\end{equation}
where $\mu_k$ expresses the rate at which the triple correlations evolve,
i.e. under viscous dissipation and nonlinear shear. It can be written as:
\begin{equation}
\mu_k=\nu k^2 + \lambda_{K}\Big(\int_0^k q^2 E(q,t) dq\Big)^{1/2} \ .
\end{equation}
Note that $\lambda_{K}$ is the only free parameter of the problem, taken equal to $0.36$ to recover the Kolmogorov constant $C_K=1.4$ for a $k^{-5/3}$ classical energy spectrum.
The expressions of $S_E(k,p,q,t)$ can be further explicited
(with the time dependency of energy spectra omitted here) as:
\begin{eqnarray}
S_E(k,p,q,t) & = &\frac{k}{pq}b\left[k^2E(q)E(p)-p^2E(q)E(k)\right]\nonumber \\
& = & S_{E_1}(k,p,q,t) + S_{E_2}(k,p,q,t) \ . \nonumber
\label{S_E}
\end{eqnarray}
Here, $S_{E_1}(k,p,q,t)$, and $S_{E_2}(k,p,q,t)$, are respectively used to denote the two terms of the extensive expression of $S_E(k,p,q,t)$.
The geometric coefficient $b(k,p,q)$
(in short, $b$ in the previous expression) is defined as:
\begin{equation}
b=\frac{p}{k}(xy+z^3)\ ,
\end{equation}
where here, $x$, $y$, $z$ are the cosines of the inner angles opposite to ${\bf k},{\bf p},{\bf q}$.

\end{document}